\documentclass[12pt,fleqn]{article}
\usepackage[utf8]{inputenc}
\usepackage{typearea}
\usepackage{amsmath,amsfonts,amssymb,amscd}
\usepackage{graphicx}
\usepackage{cite}
\usepackage{mathrsfs}
\usepackage{bbm}
\usepackage{units}
\usepackage{xcolor}
\usepackage{stmaryrd}
\usepackage{pifont}
\usepackage{hyperref}
\usepackage{cancel}
\usepackage{ytableau}
\usepackage[vcentermath]{youngtab}
\usepackage{tensor}
\usepackage{autobreak}

\newcommand{\Eqref}[1]{Eq.~\eqref{#1}}

\newcommand{\Tabref}[1]{Tab.~\ref{#1}}
\newcommand{\Secref}[1]{Sec.~\ref{#1}}
\newcommand{\Appref}[1]{App.~\ref{#1}}

\DeclareMathOperator{\im}{Im}
\DeclareMathOperator{\rank}{rank}

\newcommand{\I}{\mathrm{i}}
\newcommand{\SU}[1]{\ensuremath{\mathrm{SU}(#1)}}

\newcommand{\rep}[2][]{\ensuremath{\boldsymbol{#2}#1}}

\renewcommand{\bar}[1]{\overline{#1}}
\newcommand{\inv}{\ensuremath{\mathcal{I}}}

\newcommand{\birdtrack}[2]{\parbox{#1}{\includegraphics[width=#1]{#2}}}

\newcommand{\tyoung}[1]{\text{\tiny$\young(#1)$}} 

\newcommand{\lowrep}[1]{\mathpalette\raisesym{#1}\relax}  
\newcommand{\raisesym}[2]{\scalebox{1.2}{\raisebox{-2pt}{$#1\;\rep{#2}$}}}

\newcommand{\yi}{y_{\mathrm{i}}}
\newcommand{\yr}{y_{\mathrm{r}}}
\newcommand{\ti}{t_{\mathrm{i}}}

\newcommand{\tr}{t_{\mathrm{r}}}
\newcommand{\qi}[1]{q_{\mathrm{i}{#1}}}
\newcommand{\qr}[1]{q_{\mathrm{r}{#1}}}

\newcommand{\Inv}[1]{\ensuremath{\mathcal{I}_{{#1}}}}
\newcommand{\Jnv}[1]{\ensuremath{\mathcal{J}_{{#1}}}}

\definecolor{Gray}{gray}{0.95}
\newcommand{\bbox}[1]{\fcolorbox{gray}{Gray}{~$\displaystyle #1$~}}
\newcommand{\tbox}[1]{\fcolorbox{gray}{Gray}{~{#1}~}}

\allowdisplaybreaks[1]

\definecolor{dred}{HTML}{D95F02}
\colorlet{red}{white!15!dred}
\definecolor{darkgreen}{HTML}{1B9E77}
\definecolor{lightgray}{gray}{0.90}

\def\mytitle{Systematic construction of basis invariants \\[0.2cm] in the 2HDM}
\title{\mytitle}

\begin{document}

\pagenumbering{Alph} 
\begin{titlepage}

\vspace*{1.0cm}

\renewcommand*{\thefootnote}{\fnsymbol{footnote}}
\begin{center}
{\Large\textbf{\mytitle}}

\vspace{1.0cm}

\textbf{Andreas Trautner\footnote[1]{Email: \texttt{trautner@mpi-hd.mpg.de}}{}}
\\[5mm]
\textit{\small
Max-Planck-Institut f\"ur Kernphysik, Saupfercheckweg 1, \\
69117 Heidelberg, Germany \\
and\\
Bethe Center for Theoretical Physics und Physikalisches Institut der Universit\"at Bonn,\\
Nussallee 12, 53115 Bonn, Germany
}
\end{center}

\vspace{5mm}

\begin{abstract}
A new systematic method for the explicit construction of (basis-)invariants is introduced and employed to
construct the full ring of basis invariants of the Two-Higgs-Doublet-Model (2HDM) scalar sector.
Co- and invariant quantities are obtained by the use of hermitian projection operators.
These projection operators are constructed from Young tableaux via birdtrack diagrams
and they are used in two steps.
First, to extract basis-covariant quantities, and second, to combine the covariants in order to obtain the actual basis invariants.
The Hilbert series and Plethystic logarithm are used to find the number and structure of the complete set of generating invariants
as well as their interrelations (syzygies).
Having full control over the complete ring of (CP-even and CP-odd) basis invariants, we give a new and simple
proof of the necessary and sufficient conditions for explicit CP conservation in the 2HDM, confirming earlier results by Gunion and Haber.
The method generalizes to other models, with the only foreseeable limitation being computing power.
\end{abstract}
\end{titlepage}
\pagenumbering{arabic}

\section{Introduction}
Physical observables -- i.e.\ measurable quantities like cross sections, branching ratios, CP asymmetries \textit{etc.} -- 
must not depend on arbitrary choices of basis and notation. Consequently, observables ultimately can only depend on basis invariant quantities.

Nevertheless, in order to formulate a theory and execute computations it arguably is necessary to pick a certain basis and parametrization.
Given a theory formulated in an arbitrary basis, the following questions arise: 
\begin{itemize}
 \item How does one obtain basis invariant quantities? 
 \item How many independent basis invariant quantities exist?
 \item How are these basis invariant quantities related to physical observables?
\end{itemize}
This paper will make contact with all of these questions.
The exemplary case treated here is the scalar potential of the two-Higgs-doublet model (2HDM), see e.g.\ \cite{Branco:2011iw}. For this case, we will be able to give definite answers 
to questions one and two. Moreover, physical matters will be touched with respect to the violation of the combined charge-conjugation and parity (CP) symmetry.

Physical questions can be obscured by the presence of large basis redundancies, in which case
an invariant formulation offers a clear benefit \cite{Santamaria:1993ah}.
In particular -- but certainly not limited to it -- this is true for detecting the (non-)conservation of CP symmetry,
which is commonly plagued by spurious phases~\cite{Branco:1999fs}. Therefore, perhaps the most prominent example for a basis invariant measure is the 
well-known Jarlskog invariant, which detects CP non-conservation in the Standard Model (SM) \cite{Jarlskog:1985ht}. 

While the non-vanishing of any CP-odd basis invariant unambiguously signals CP violation, sufficient conditions for CP conservation
are much harder to find. That is because in order to formulate sufficient conditions one has to know ``when to stop looking'' 
for new independent invariants, which is of course at the heart of our second question above.
In this light, it is no surprise that basis invariant necessary conditions for CP conservation have been formulated for theories 
with extended fermion \cite{Bernabeu:1986fc,Branco:1986gr} and/or scalar sectors \cite{Botella:1994cs,Lavoura:1994fv},
while sufficient conditions for CP conservation, besides for the SM, are only known for the 2HDM \cite{Gunion:2005ja} (see also \cite{Branco:2005em,Ivanov:2005hg,Nishi:2006tg}) 
and under limiting assumptions also for the three and N-HDM \cite{Nishi:2006tg}, as well as for certain supersymmetric models \cite{Lebedev:2002wq,Dreiner:2007yz}. 
Lately, also basis invariant necessary and sufficient conditions for the physically distinct order $4$ CP transformation of the 3HDM 
\cite{Ivanov:2015mwl} have been formulated \cite{Haber:2018iwr,Ivanov:2018ime}.

Basis invariant methods have also been used to grasp physical aspects other than CP.
For example, to investigate quark and lepton mixing in the SM and extensions \cite{Jenkins:2007ip,Jenkins:2009dy,Hanany:2010vu},
as well as to express physical observables of the 2HDM \cite{Davidson:2005cw,Haber:2006ue}, and more recently 
\cite{Grzadkowski:2016szj,Bento:2017eti,Ogreid:2018bjq}. 
As an additional benefit, a basis invariant parametrization simplifies the analysis of renormalization group equations (RGE) and RGE running,
both for SM fermions \cite{Feldmann:2015nia, Chiu:2015ega, Chiu:2016qra} as well as for extended scalar sectors \cite{Herren:2017uxn, Bednyakov:2018cmx, Bijnens:2018rqw},
and so the question of how to construct basis invariants continues to be of interest \cite{Varzielas:2016zjc,Berger:2018dxg}.
However, it remains in general an open question how a theory can be formulated
solely in terms of basis invariant quantities.

In a group theoretical sense, basis invariants are objects which do not transform under the action of the group of basis changes.
This implies that basis invariants form a ring, in the algebraic sense, and 
the question ``when to stop looking'' for new invariants turns out to be a mathematical exercise of invariant theory (see e.g.\ \cite{stanley1979, Sturmfels:2008}).
The classical way of dealing with invariant rings is via their generating function, the so-called Hilbert-Poincar\'e series (HS).
A more modern tool in ring theory is the Plethystic logarithm (PL), introduced in \cite{1994dg.ga.....8003G} and further discussed in \cite{Labastida:2001ts}.
The methodology of the HS and PL has been put forward for physics applications in \cite{Benvenuti:2006qr,Feng:2007ur},
and has subsequently been applied to a plethora of formal questions, see e.g.\ \cite{Noma:2006pe,Butti:2007jv,Gray:2008yu,Hanany:2008kn,Hanany:2008sb,Hanany:2014dia,Bourget:2017tmt}.
Important phenomenological applications of the HS method are the characterization of quark and lepton invariants in and beyond the SM \cite{Jenkins:2009dy,Hanany:2010vu},
as well as the general construction of complete bases of gauge invariant operators in effective field theories \cite{Henning:2015daa,Henning:2015alf,Henning:2017fpj}.
We recommend \cite{Lehman:2015via} as a first read on HS in the particle physics context. 

In this paper, we will use the HS and PL to find the number of independent basis invariants, a generating set
of basis invariants, as well as the structure of interrelations between basis invariant (syzygies) -- all for the 2HDM. 

The main original aspect of the present paper is our novel way of explicitly constructing the basis invariants in a systematic way.
Of course, the construction of invariants from covariant objects is, in principle, a solved group theoretical problem for which even powerful computer codes exist 
(see e.g.\ \cite{Fonseca:2011sy}). 
However, available methods quickly become unmanageable if it comes to construction of singlets from high-rank tensors or disentangling individual contributions
of, in principle, independent singlets. Our approach is to use hermitian projection operators 
\cite{Keppeler:2013yla} (see also \cite{Alckock-Zeilinger:2016bss,Alcock-Zeilinger:2016cva,Alcock-Zeilinger:2016sxc}),
which can conveniently be constructed from Young tableaux via birdtrack diagrams \cite{Cvitanovic:1976am,Cvitanovic:2008zz}.
As a pedagogical introduction to this we recommend \cite{Keppeler:2017kwt}.
These operators project arbitrary rank tensors onto their contained orthogonal trivial singlets, implying that they give rise to the shortest possible invariants by construction.
Ultimately, it is the liaison of group theoretical and algebraic techniques that merges in a new and powerful systematic way for the construction
of basis invariants. The thereby constructed invariants are short and their relations transparent, as demonstrated by our new and 
simple proof of necessary and sufficient conditions for explicit CP conservation in the 2HDM.

The outline of the paper is as follows. In the subsequent section we will give a brief synopsis of indispensable terms and concepts. 
In \Secref{sec:buildingblocks} we turn to the 2HDM, for which we construct the elementary building blocks 
of all invariants from the coupling tensors of the Lagrangian.
The CP transformation behavior of the building blocks and higher-order invariants is derived in \Secref{sec:CPproperties}.
Using the building blocks as input, we proceed to construct the Hilbert series and Plethystic logarithm of the ring in \Secref{sec:construction},
thereby gaining information on the number and structure of the invariants and their syzygies.
The explicit construction of the invariants is performed in \Secref{sec:primary} and \Secref{sec:GeneratingSet},
and a generally applicable strategy for the construction of syzygies is outlined in \Secref{sec:syzygies}.
We then give a simple derivation of the necessary and sufficient conditions for explicit CP conservation
in the 2HDM in \Secref{sec:ConditionsforCPC}.
Finally, \Secref{sec:Hiroanaka} contains some comments on the construction of a Hironaka decomposition of the 2HDM ring, after which we conclude.

\section{Synopsis of jargon}
Before starting the actual discussion of this paper let us introduce the technical terms used.
All parameters of a theory do transform in some way or another  
under necessarily unphysical changes of basis. 
Thus, it is in general possible to form combinations of parameters which are unaffected by 
\textit{all} possible basis transformations: basis invariants (or short invariants in the following).
Since any sum or product of invariants is an invariant itself, 
the invariants form a \textbf{ring} in the mathematical sense.\footnote{%
Most likely the invariants form even more than a ring, since one may 
also envisage more involved operations on them.}
An immediate question then is: how many invariant quantities are needed, in general, 
in order to be able to cover the space of \textit{all} possible invariant quantities of a given model.

Some remarks are in order concerning the use of ``generating set'', ``generators'', ``independent'' and so forth 
since most of the physics literature lacks scrutiny in this point.
By ``independent'' we always mean \textit{algebraically} independent.
To be clear, an invariant, say $\inv_1$, is \textbf{algebraically dependent} on a set of invariants, say $\inv_{2,3,..}$ , if and only if it is possible to find a polynomial $P$ such that
\begin{equation}\label{eq:independence}
 P\left(\inv_1,\inv_2,\inv_3,\dots\right)=0\;.
\end{equation}
If such a polynomial does not exist, $\inv_1$ is called algebraically independent of $\inv_{2,3,..}$.
The maximal number of algebraically independent invariants is equal to the number of physical parameters of a theory
in the usual sense. 
Not surprisingly, this is the number of parameters which remains after all possible basis changes have been used 
to absorb parameters, i.e.\ set as many of them to zero as possible.
Thus, having found a full set of algebraically independent invariants, the physical content of a theory is fully specified.
Nevertheless, for many applications it makes sense to go beyond the set of algebraically independent invariants.
This is the case, particularly because a relation of the kind \eqref{eq:independence} does not guarantee that 
we can solve for an arbitrary invariant. For this reason, it makes sense to discuss a generating set of a ring,
which consists of all invariants that \textit{cannot} be written as a polynomial of other invariants,%
\begin{equation}
 \inv_i\neq P\left(\inv_j,\dots\right)\;.
\end{equation}
More intuitively perhaps, the \textbf{generating set} of invariants is a set such that \textit{all} invariants in the ring can
be written as polynomial in the generators,%
\begin{equation}\label{eq:generalI}
\inv=P\left(\inv_1,\inv_2\dots\right)\;.
\end{equation}
A classic result is that the generating set of a ring of invariants
has finite size (``the ring is finitely generated'') 
if the underlying group of symmetry transformations
(in our case the group of basis transformations) is reductive \cite{Noether:1916,Hochster:1974a, Hochster:1974b}.\footnote{%
All finite groups and all semi-simple Lie groups are reductive.}
General formulae for the number of invariants in the generating set are not available, but there exist bounds \cite{Knop1987,KNOP198940} 
(see also \cite{Jenkins:2009dy} for applications).

The full set of generators of a ring is, generally, algebraically dependent (otherwise the ring is called free).
Algebraic relations among invariants are called \textbf{syzygies}.

\section{Construction of the building blocks}
\label{sec:buildingblocks}
The most general 2HDM scalar potential can be written as
\begin{equation}\label{eq:Potential}
 V=\Phi^\dagger_a\,Y\indices{^a_b}\,\Phi^b+\Phi^\dagger_a\,\Phi^\dagger_b\,Z\indices{^{ab}_{cd}}\,\Phi^c\,\Phi^d\;,
\end{equation}
where $\Phi^a$ are hyper charge-one scalar doublets of the $\SU2_{\mathrm{L}}$ gauge symmetry and $a,b,c,d=1,2$ are indices in the \SU2 space of Higgs-flavor.
We use upper and lower indices to distinguish fields transforming as $\rep{2}$ and $\rep{\bar{2}}$ under \SU2 Higgs-flavor basis changes.

Due to hermiticity of the Lagrangian, the coupling tensors satisfy
\begin{equation}\label{eq:hermiticity}
 Y\indices{^a_b}=(Y\indices{^b_a})^*\;,\qquad Z\indices{^{ab}_{cd}}=(Z\indices{^{cd}_{ab}})^*\;.
\end{equation}
Because of $\SU2_{\mathrm{L}}$ invariance the quartic couplings fulfill in addition
\begin{equation}\label{eq:SymZ}
 Z\indices{^{ab}_{cd}}=Z\indices{^{ba}_{dc}}\;.
\end{equation}
Due to these constraints, not all entries of the coupling tensors are independent.
$Y$ and $Z$ have $4$ and $10$ independent real entries, respectively.
Utilizing all possible basis changes to absorb parameters (i.e.\ set them to zero) one can show that the number of physical parameters is $11$.
We will, however, keep working in a general basis.

Our first goal is to construct linear combinations of the entries of $Y$ and $Z$ which transform 
in irreducible representations under basis changes.
In particular we will see that $Y$ and $Z$ decompose into 
\begin{align}\label{eq:YZdecomposition}\nonumber
 Y~&\mathrel{\widehat=}~\rep{1}\oplus\rep{3}\;,\\
 Z~&\mathrel{\widehat=}~\rep{1}\oplus\rep{1}\oplus\rep{3}\oplus\rep{5}\;.
\end{align}
There are three ``trivial'' basis invariants which arise as \textit{linear} combinations of the potential parameters.
Linearly independent of these, there are other linear combinations of entries of $Y$ and $Z$ which transform covariantly under basis changes, and they will be our \textbf{building blocks} 
for the construction of \textit{non-linear} higher-order invariants below.

Let us be very explicit in deriving \eqref{eq:YZdecomposition}.
The undertaken steps may seem like an overkill to advanced readers, 
but it seems necessary to recall these details before entering the derivation of higher-order invariants below.

As we want to make use of Young tableaux, it makes sense to convert all indices to one type (upper or lower).
To do this, we use the fully anti-symmetric Levi-Civita tensor with the convention $\varepsilon^{12}=-\varepsilon^{21}=-\varepsilon_{12}=\varepsilon_{21}=1$ to raise or lower indices,
thereby defining the objects
\begin{equation}
 Y^{ab}:=\varepsilon^{bc}\,Y\indices{^a_c}\;,\qquad\text{and}\qquad Z^{ab,cd}:=\varepsilon^{ce}\,\varepsilon^{df}\,Z\indices{^{ab}_{ef}}\;.
\end{equation}
Assigning a box to each index in the usual way, we can decompose $Y$ and $Z$ into their covariantly transforming irreducible components by the standard procedure,\footnote{%
Due to the constraints \eqref{eq:hermiticity} and \eqref{eq:SymZ}, only one of the three triplets \rep[_{(1,2,3)}]{3} in the decomposition of $Z$ is independent, such that
Eq.\ \eqref{eq:YZdecomposition} holds. We will show this explicitly around Eq.\ \eqref{eq:triplets}.}
\begin{align}
\ytableausetup{centertableaux,boxsize=1.2em}
&Y:&
\begin{ytableau}
 a
\end{ytableau}_{\lowrep{2}}
\otimes
\begin{ytableau}
 b
\end{ytableau}_{\lowrep{2}}
~=~&
\begin{ytableau}
 a \\ b
\end{ytableau}_{\lowrep{1}}
\oplus
\begin{ytableau}
 a & b
\end{ytableau}_{\lowrep{3}}\;,& \\\nonumber
&Z:&
\begin{ytableau}
 a
\end{ytableau}_{\lowrep{2}}
\otimes
\begin{ytableau}
 b
\end{ytableau}_{\lowrep{2}}
\otimes
\begin{ytableau}
 c
\end{ytableau}_{\lowrep{2}}
\otimes
\begin{ytableau}
 d
\end{ytableau}_{\lowrep{2}}
~=~&
\begin{ytableau}
 a & b \\ 
 c & d
\end{ytableau}_{\lowrep{1}_{{}_{(1)}}}
\oplus
\begin{ytableau}
 a & c \\
 b & d 
\end{ytableau}_{\lowrep{1}_{{}_{(2)}}}\,\oplus & \\\nonumber
&&&
\begin{ytableau}
 a & b & c \\
 d  
\end{ytableau}_{\hspace{-28pt}\lowrep{3}_{{}_{(1)}}}\hspace{20pt}
\oplus
\begin{ytableau}
 a & b & d \\
 c  
\end{ytableau}_{\hspace{-28pt}\lowrep{3}_{{}_{(2)}}}\hspace{20pt}
\oplus
\begin{ytableau}
 a & c & d \\
 b  
\end{ytableau}_{\hspace{-28pt}\lowrep{3}_{{}_{(3)}}}\,\hspace{20pt}\oplus & \\
&&&
\begin{ytableau}
 a & b & c & d  
\end{ytableau}_{\lowrep{5}}
\;.&
\label{eq:YZdecompositionYT}
\end{align}
Recall that writing boxes in a line of the Young tableaux means symmetrization of the respective indices, while boxes in the same column are to be anti-symmetrized. 
The order of indices has to be obeyed. The (anti-)symmetrization of indices can be performed by projection operators.
Thus, in their very essence, Young tableaux correspond to projection operators. These projection operators can always be made hermitian \cite{Keppeler:2013yla}.
Acting with these projection operators on a corresponding tensor projects out the respective covariantly transforming components of the tensor. 
Birdtracks \cite{Cvitanovic:1976am,Cvitanovic:2008zz} can be used to construct the projection operators by hand. 
A pedagogical introduction to this can be found in \cite{Keppeler:2017kwt}. We will only recall the very most relevant features along the way.

The simplest projection operators are the total (anti-)symmetrizers of two indices denoted by
\begin{align}
  P_{_{\tyoung{a,b}}}:=\birdtrack{16ex}{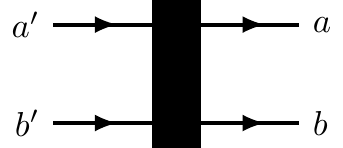}&~\equiv~\frac{1}{2!}\left(
  \birdtrack{9ex}{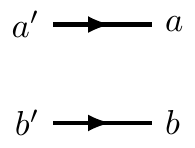} - \birdtrack{9ex}{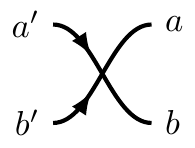}\right)\;,\\[0.3cm]
  P_{_{\tyoung{ab}}}:=\birdtrack{16ex}{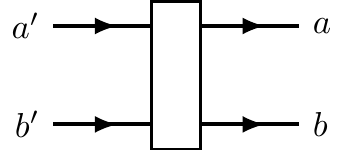}&~\equiv~\frac{1}{2!}\left(
  \birdtrack{9ex}{Birdtracks/Prop2} + \birdtrack{9ex}{Birdtracks/Prop2perm12}\right)\;.
\end{align}
Indices on incoming(outgoing) lines correspond to upper(lower) indices, and if two indices are directly connected by a line they are meant to be set equal by contraction with a Kronecker delta. 
Using only these and larger (anti-)symmetrization operators consecutively, it is possible to diagrammatically construct projection operators for every Young tableaux. 

Let us also introduce a diagrammatic representation of the Levi-Civita symbol
\begin{subequations}
\begin{align}
 \varepsilon^{ab}~\equiv&~\birdtrack{14ex}{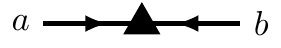}=-\birdtrack{14ex}{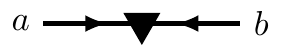}\;,\\
 \varepsilon_{ab}~\equiv&~\birdtrack{14ex}{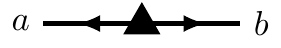}=-\birdtrack{14ex}{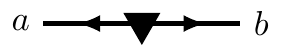}\;,
\end{align}
\end{subequations}
with the usual identities
\begin{equation}
 \birdtrack{15ex}{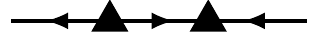}~=~\birdtrack{15ex}{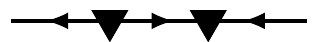}~=~\birdtrack{5ex}{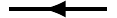}\;, \quad etc.\;.
\end{equation}
It is very important to note that for indices in the fundamental of $\SU{N}$ we can anti-symmetrize at most $N$ indices, otherwise the result is trivially zero.
Diagrammatically, this 
gives rise to a \textit{factorization rule} for anti-symmetrization operators with $N$ indices. Here, for $N=2$:\footnote{%
Spelled out as $\delta\indices{^{a'}_a}\delta\indices{^{b'}_b}-\delta\indices{^{a'}_b}\delta\indices{^{b'}_a}=-\varepsilon^{a'b'}\varepsilon_{ab}$,
this is nothing but the well-known Schouten identity for \SU2 with two indices lowered (see e.g.\ \cite{Dreiner:2008tw}).
The crucial point here is really that this leads to (in some cases only partly) factorization of this and much larger projection operators (see below),
also generalizing to \SU{N}.}
\begin{equation}\label{eq:factorization}
 P_{_{\tyoung{a,b}}}~=~\birdtrack{16ex}{Birdtracks/As2ab}~=~\frac{1}{2}\left(\birdtrack{18ex}{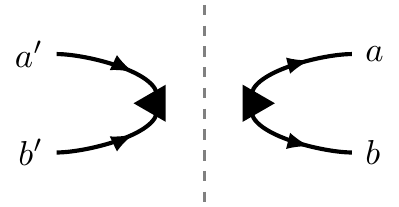}\right)~=~-\frac{1}{2}\varepsilon^{a'b'}\varepsilon_{ab}\;.
\end{equation}
Finally, let us also introduce a diagrammatic notation for $Y$ and $Z$,
\begin{align}
 Y^{ab} &~=~\birdtrack{14.4ex}{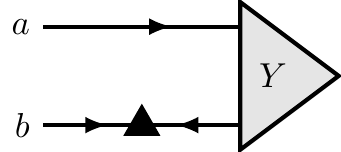}~=:~\birdtrack{10.7ex}{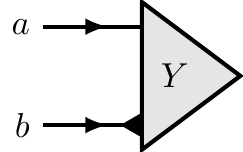}\;,\qquad\text{and} \\[0.3cm]
 Z^{ab,cd}&~=~\birdtrack{18.9ex}{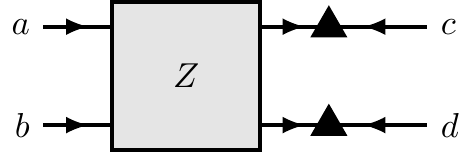}~=:~\birdtrack{15ex}{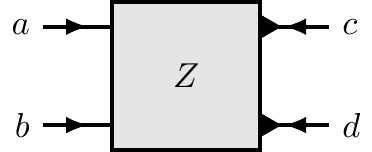}\;.
\end{align}

To find the covariantly transforming components of the tensor $Y$, we simply act on it with the corresponding projection operators,
\begin{equation}
 Y_{\rep{1}}^{a'b'}=[P_{_{\tyoung{a,b}}}]\indices{^{a'b'}_{ab}}\,Y^{ab}\;,\qquad\text{and}\qquad Y_{\rep{3}}^{a'b'}=[P_{_{\tyoung{ab}}}]\indices{^{a'b'}_{ab}}\,Y^{ab}\;.
\end{equation}
Diagrammatically the first projection reads
\begin{equation}
 Y_{\rep{1}}^{a'b'}~=~\frac{1}{2}\birdtrack{20ex}{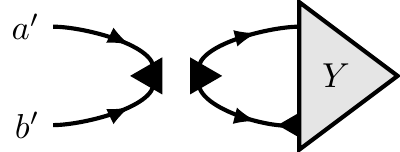}~=~\frac{1}{2}\birdtrack{20ex}{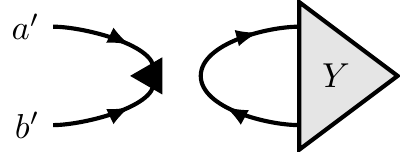}~=~-\frac{1}{2}\varepsilon^{a'b'}\,Y\indices{^a_a}\;.
\end{equation}
Note how due to \textit{factorization} of the projection operator a \textit{bubble diagram} is \textit{nucleated}.
Clearly, $Y_{\rep{1}}^{a'b'}$ is a basis invariant because $\varepsilon^{a'b'}$ is invariant under any \SU2 rotation while $Y\indices{^a_a}\!$ is fully contracted.
We are just interested in the non-trivial essence of the invariant. Hence, the global prefactors 
as well as the remaining $\varepsilon$-tensor are irrelevant and we will just drop them. 
The first basis invariant hence is
\begin{equation}
 Y_{\rep{1}}:=\birdtrack{10ex}{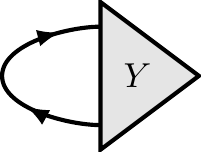}=Y\indices{^a_a}\;.
\end{equation}

Factorization does not take place for the second projection with the symmetrizer, which results in 
\begin{equation}
 Y_{\rep{3}}^{a'b'}:=\birdtrack{20ex}{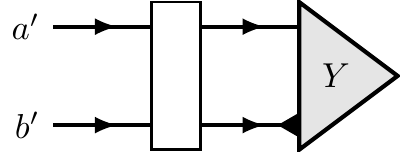}=\frac12\left(Y^{a'b'}+Y^{b'a'}\right)\;.
\end{equation}

The hermitian projection operators needed to decompose $Z$ are more involved. They can be constructed following the rules of \cite{Alckock-Zeilinger:2016bss,Alcock-Zeilinger:2016cva,Alcock-Zeilinger:2016sxc}.\footnote{
While the rules in \cite{Alckock-Zeilinger:2016bss,Alcock-Zeilinger:2016cva,Alcock-Zeilinger:2016sxc} are derived with mathematical rigor,
following them to construct projection operators in practice can be tedious. 
We remark here that there is a set of simple and very intuitive rules to construct these (and much larger) projection operators from scratch.
We plan to communicate these rules in the future.}
The projectors for the singlets read
\begin{equation}\label{eq:P4S}
 P_{_{\tyoung{ab,cd}}}=\frac43\;\birdtrack{10ex}{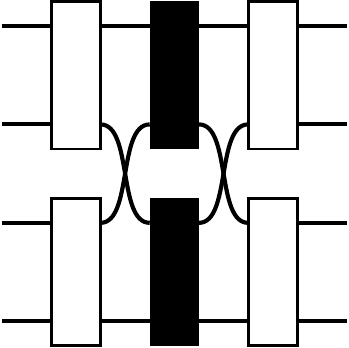}\;,\quad\text{and}\qquad P_{_{\tyoung{ac,bd}}}=\frac43\;\birdtrack{10ex}{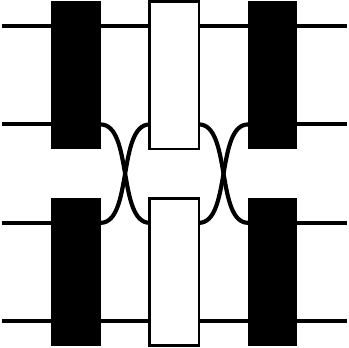}\;,
\end{equation}
while the triplet projection operators are given by
\begin{equation}\label{eq:TripletProjectors}
P_{_{\tyoung{abc,d}}}=\frac32\;\birdtrack{10ex}{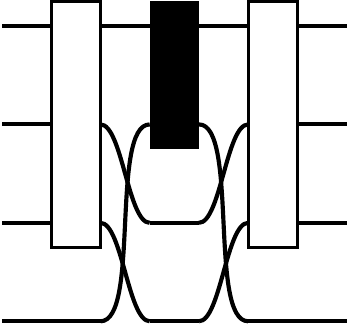}\;, \quad P_{_{\tyoung{acd,b}}}=\frac32\;\birdtrack{10ex}{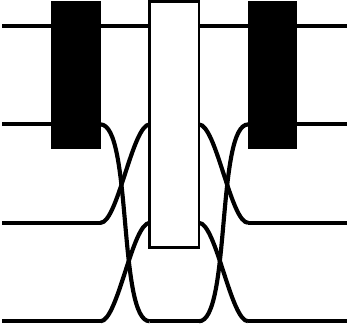}\;, \quad P_{_{\tyoung{abd,c}}}=2\;\birdtrack{14.6ex}{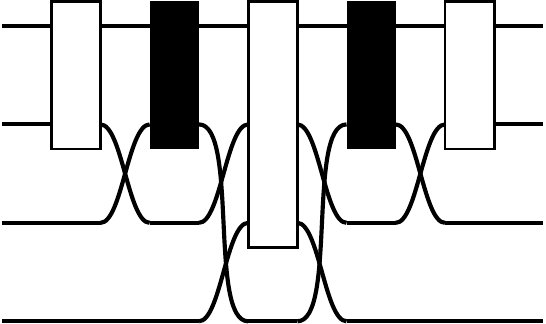}\;.
\end{equation}
Here and in the following we will drop arrows and indices on birdtracks whenever restoring them works in the obvious way. 
The last remaining projection operator is $P_{\rep{5}}$ to project on \,\raisebox{2pt}{\scalebox{0.7}{$\begin{ytableau}a & b & c & d\end{ytableau}$}}\,. 
This is simply the total symmetrizer of four indices, and so we do not display it.

Again, note how due to the factorization rule \eqref{eq:factorization}, 
both of the singlet projection operators in \eqref{eq:P4S} \textit{factorize}. Upon projection this \textit{nucleates a vacuum bubble diagram} which transforms as a trivial singlet.
Acting with the operators of \Eqref{eq:P4S} on $Z$, the two singlets result as
\begin{align}
Z_{\rep{1}_{(1)}}&:=\birdtrack{17ex}{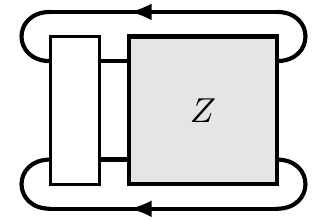}=\frac12\left(Z\indices{^{ab}_{ab}}+Z\indices{^{ab}_{ba}}\right)\;, \quad\text{and} \\[0.3cm]
Z_{\rep{1}_{(2)}}&:=\birdtrack{18ex}{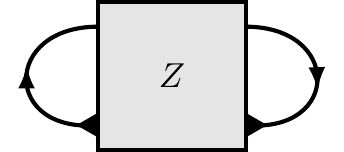}=\varepsilon_{ab}\varepsilon^{cd}Z\indices{^{ab}_{cd}}\;.
\end{align}

Next we construct the triplet building blocks of $Z$.
However, before applying projection operators on $Z$ to extract the triplet irreps, observe that due to the symmetry in \Eqref{eq:SymZ}, $Z$ identically decomposes into
\begin{equation}\label{eq:Zdecomposition}
 \birdtrack{19.66ex}{Birdtracks/Z}=\birdtrack{27.4ex}{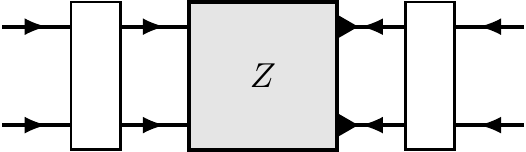}+\birdtrack{27.4ex}{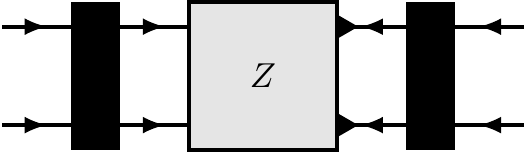}\;.
\end{equation}
The second term is immediately identified as the trivial singlet $Z_{\rep{1}_{(2)}}$ by using the factorization rule \eqref{eq:factorization}.
Hence, it will not contribute to the triplet irreps. 
Using this decomposition it is also easy to see that the triplet $\rep[_{(2)}]{3}$ vanishes identically: the corresponding projection operator (in the middle of \eqref{eq:TripletProjectors}) 
anti-symmetrizes the first two indices and symmetrizes the last two indices. Thus, acting with it on $Z$ (using the decomposition \eqref{eq:Zdecomposition})
it is unavoidable that symmetrizers and anti-symmetrizers get mutually connected -- which always annihilates any contribution.
The remaining two triplets, $\rep[_{(1)}]{3}$ and $\rep[_{(3)}]{3}$ of \Eqref{eq:YZdecompositionYT},  are degenerate. 
This can be shown via a straightforward direct computation, or alternatively diagrammatically. We find
\begin{equation}\label{eq:triplets}
 \frac32\,Z_{\rep[_{(1)}]{3}}^{ab}=\,Z_{\rep[_{(3)}]{3}}^{ab}=\birdtrack{28.3ex}{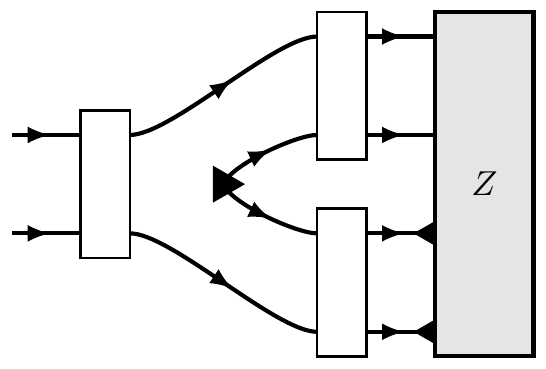}=:Z_{\rep{3}}^{ab}\;,
\end{equation}
where we have deformed the box of $Z$ to make the result more pleasant to the eye.
Spelling out $Z_{\rep{3}}^{ab}$ explicitly would already be cumbersome, while diagrammatically it can be represented in a compact way.

Finally, the five-plet $Z_{\rep{5}}$ is simply given by the total symmetrization of all indices of $Z$ and so we do not show this.

To conclude their construction, we explicitly state the obtained building blocks in terms of the components of $Y$ and $Z$.
Denoting their components as $\left[Y\right]\indices{^a_b}=y_{ab}$ and $\left[Z\right]\indices{^{ab}_{cd}}=z_{abcd}$, and dropping the irrelevant global prefactors 
of all covariants, one finds
\begin{subequations}\label{eq:buildingblocks}
\begin{align}
 Y_{\rep[_{\phantom{{(1)}}}]{1}}&~=~y_{11}+y_{22}\;,& \\
 Z_{\rep[_{(1)}]{1}}&~=~z_{1111}+z_{1212}+z_{1221}+z_{2222}\;,& \\
 Z_{\rep[_{(2)}]{1}}&~=~z_{1212}-z_{1221}\;,& \\
 Y_{\rep{3}}^{ab}&~=~\left(
 \begin{array}{cc}
 y_{12} & \frac{1}{2} \left(y_{22}-y_{11}\right) \\
 \frac{1}{2} \left(y_{22}-y_{11}\right) & -y_{12}^* \\
 \end{array}
 \right)\;,& \\ 
 Z_{\rep{3}}^{ab}&~=~\left(
 \begin{array}{cc}
 \phantom{\frac12}\,z_{1112}+z_{1222} & \frac{1}{2} \left(z_{2222}-z_{1111}\right) \\
 \frac{1}{2} \left(z_{2222}-z_{1111}\right) & \,-\left(z_{1112}+z_{1222}\right)^* \\
 \end{array}
 \right)\;,&\\ 
 Z_{\rep{5}}^{abcd}&~=~\left(
\begin{array}{cc}
 \left(
\begin{array}{cc}
 \phantom{-}\zeta_1 & \phantom{-}\zeta_2 \\
 \phantom{-}\zeta_2 & \phantom{-}\zeta_3 \\
\end{array}
\right) & \left(
\begin{array}{cc}
 \phantom{-}\zeta_2 & \phantom{-}\zeta_3  \\
 \phantom{-}\zeta_3 & -\zeta_2^* \\
\end{array}
\right) \\[0.3cm]
 \left(
\begin{array}{cc}
 \phantom{-}\zeta_2 & \phantom{-}\zeta_3  \\
 \phantom{-}\zeta_3 & -\zeta_2^* \\
\end{array}
\right) & \left(
\begin{array}{cc}
 \phantom{-}\zeta_3 & -\zeta_2^*  \\
 -\zeta_2^* & \phantom{-}\zeta_1^* \\
\end{array}
\right) \\
\end{array}
\right)
 \;.&
\end{align}
\end{subequations}
Here, $\left\{y_{11},y_{22}, z_{1111}, z_{1212}, z_{1221}, z_{2222} \right\}\in\mathbb{R}$, $\left\{y_{12}, z_{1122}, z_{1222}, z_{1112} \right\}\in\mathbb{C}$, and
\begin{subequations}
\begin{align}
 \zeta_1&:=z_{1122}\;,& \\
 \zeta_2&:= \frac{1}{2} \left(z_{1222}-z_{1112}\right)\;,& \\
 \zeta_3&:=\frac{1}{6} \left(z_{1111}-2 z_{1212}-2 z_{1221}+z_{2222}\right)\;.&
\end{align}
\end{subequations}
This explicitly shows how the $14$ independent parameters of $Y$ and $Z$ combine to form the irreps of \Eqref{eq:YZdecomposition}.
For completeness, we have also stated the building blocks in the conventional parametrization of the 2HDM scalar potential in \Appref{app:LambdaNotation}.

\section{CP properties of the building blocks}
\label{sec:CPproperties}
Having the complete set of building blocks at hand, let us study their transformation behavior under general CP symmetries.
Under a general CP transformation the scalars transform as\footnote{%
We focus here on the transformation behavior in the \SU2 Higgs-flavor space, hence suppress the transformation with respect to spacetime and internal gauge symmetries.}
\begin{align}
 &\Phi^a\mapsto\Phi_b^*\,[U^{\mathrm{T}}]\indices{^b_a}\;,&  &\Phi^*_a\mapsto [U^*]\indices{^a_b}\,\Phi^b\;.&
\end{align}
Applied to \eqref{eq:Potential} we see that this transformation can equivalently be described by a mapping of parameters
\begin{align}
 Y\indices{^a_b}&\mapsto \left[U^{\mathrm{T}}\,Y^{\mathrm{T}}\,U^*\right]\indices{^a_b}\;,&\\
 Z\indices{^{ab}_{cd}}&\mapsto [U^{\mathrm{T}}]\indices{^a_{a'}}\,[U^{\mathrm{T}}]\indices{^b_{b'}}\,[Z^{\mathrm{T}}]\indices{^{a'b'}_{c'd'}}\,[U^*]\indices{^{c'}_c}\,[U^*]\indices{^{d'}_d}\;.&
\end{align}
Assuming $U$ to be symmetric\footnote{%
This implies that we restrict ourselves here to CP transformation of order two, see e.g.\ \cite{Ecker:1987qp}.
For higher-order CP transformations one has to carry through the matrix $U$, but the result is the same.} 
one can always choose a basis in flavor space in which $U\propto\mathbbm{1}$.
Assuming for a minute that we had transformed the theory to such a basis, the transformation of the coupling tensors simplifies to
\begin{align}
 Y\indices{^a_b}&\mapsto Y\indices{^b_a}=(Y\indices{^a_b})^*\;,&\\
 Z\indices{^{ab}_{cd}}&\mapsto Z\indices{^{cd}_{ab}}=(Z\indices{^{ab}_{cd}})^*\;,&
\end{align}
where we have used hermiticity, cf.\ \eqref{eq:hermiticity}, for the last equalities. 
For the building blocks this implies the transformation
\begin{align}\notag
 &Y_{\rep{1}}^{\phantom{ab}}\mapsto Y_{\rep{1}}\;,&  &Z_{\rep[_{(1)}]{1}}\mapsto Z_{\rep[_{(1)}]{1}}\;,& &Z_{\rep[_{(2)}]{1}}^{\phantom{abcd}}\mapsto Z_{\rep[_{(2)}]{1}}\;,& \\
 &Y_{\rep{3}}^{ab}\mapsto -({Y_{\rep{3}}})_{ab}\;,&  &Z_{\rep[_{\phantom{(1)}}]{3}}^{ab}\mapsto -({Z_{\rep{3}}})_{ab}\;,&  &Z_{\rep{5}}^{abcd}\mapsto (Z_{\rep{5}})_{abcd}\;.&
\end{align}
It follows that basis invariants (for which all indices have to be contracted) can at most transform with a sign under a CP transformation of order two.
Furthermore, we immediately recognize a simple rule to distinguish CP-even and CP-odd basis invariants:
\begin{center}
\tbox{\parbox{\dimexpr\textwidth-30\fboxsep}{\centering 
A basis invariant is CP $\left\{\begin{array}{cc}\text{even} \\ \text{odd}\end{array} \right\}$ \textit{iff} it contains an \\ $\left\{\begin{array}{cc}\text{even} \\ \text{odd}\end{array} \right\}$ number of 
triplet building blocks ($Y_{\rep{3}}$, $Z_{\rep{3}}$).
}}
\end{center}

\section{Construction of higher-order invariants}
\label{sec:invariants}
\ytableausetup{smalltableaux}
\subsection{Number and structure of the invariants}

We have identified all possible linear invariants. 
Our goal is now to construct \textit{non-linear} (in the potential parameters) higher-order invariants out of the covariantly transforming building blocks $Y_{\rep{3}}$, $Z_{\rep{3}}$, and $Z_{\rep{5}}$.
To construct these invariants explicitly, we will once again make use of hermitian Young projection operators.
However, before we do this we first want to determine the size and structure (in terms of the building blocks) of the generating set of invariants.
Even though it is conceivable that even this step can be performed diagrammatically, this seems like a rather tedious way to progress. 
We will ease this step by using the (multi-graded) Hilbert series \cite{Benvenuti:2006qr,Feng:2007ur,Gray:2008yu} (see \cite{Lehman:2015via} for an accessible introduction and \cite{Jenkins:2009dy} for 
many examples in the particle physics context).
The Hilbert series (HS) together with the Plethystic logarithm (PL) will allow us to fully characterize the ring of basis invariants and, furthermore, reveal the structure of 
all sought invariants. This input then will be merged with our diagrammatic approach to finally construct all required invariants explicitly.

The linear invariants are irrelevant for the construction of non-trivial higher-order invariants.
We will, therefore, focus only on the non-trivially transforming building blocks in this section and add the linear invariants back in later.

We will first derive the (multi-graded) HS and PL and then discuss their information content. 
To ease the notation we define the symbols%
\begin{center}
 $\bbox{y~\mathrel{\widehat=}~Y_{\rep{3}}\;,\quad t~\mathrel{\widehat=}~Z_{\rep{3}}\;, \quad\text{and}\quad q~\mathrel{\widehat=}~Z_{\rep{5}}\;.}$
\end{center}
We will need the character polynomials $\chi_{\rep{r}}(z)$ for the relevant \SU2 irreps \rep{r}, which are given by (see e.g.\ \cite[App.\ A.2]{Lehman:2015via})
\begin{subequations}
\begin{align}
\chi_{\rep{3}}(z)&=z^2+1+\frac{1}{z^2}\;,& \\
\chi_{\rep{5}}(z)&=z^4+z^2+1+\frac{1}{z^2}+\frac{1}{z^4}\;,&
\end{align}
\end{subequations}
as well as the plethystic exponential (PE), which is defined as (see e.g.\ \cite{Benvenuti:2006qr,Feng:2007ur})
\begin{equation}
 \mathrm{PE}\left[z,x,\rep{r}\right]:=\mathrm{exp}\left(\sum\limits_{k=1}^{\infty}\frac{x^k\,\chi_{\rep{r}}(z^k)}{k}\right)\;.
\end{equation}
Using the token variables $q$, $y$, and $t$ as defined above, the multi-graded HS is computed as (see e.g.\ \cite{Hanany:2008sb} for the integral measure)
\begin{equation}
\mathfrak{H}(q,y,t)= \frac{1}{2\pi\I}\oint\limits_{|z|=1}\frac{\mathrm{d}z}{z}\left(1-z^2\right) \mathrm{PE}\left[z,q,\rep{5}\right]\,\mathrm{PE}\left[z,y,\rep{3}\right]\,\mathrm{PE}\left[z,t,\rep{3}\right]\;.
\end{equation}
Performing the integration via the residue theorem we find
\begin{equation}\label{eq:MGHS}
\mathfrak{H}(q,y,t)=\frac{N\left(q,y,t\right)}{D\left(q,y,t\right)}\;,
\end{equation}
with the numerator
\begin{equation}
\begin{split}
N\left(q,y,t\right)=\,&1+q t y+q^2 t y+q t^2 y+q t y^2+q^2 t^2 y+q^2 t y^2\\
&+q^3 t^3+q^3 t^2 y+q^3 t y^2+q^3 y^3\\
&-q^3 t^4 y-q^3 t^3 y^2-q^3 t^2 y^3-q^3 t y^4-q^4 t^3 y^2-q^4 t^2 y^3\\
&-q^5 t^3 y^2-q^5 t^2 y^3-q^4 t^3 y^3-q^5 t^3 y^3-q^6 t^4 y^4\;,
\end{split}
\end{equation}
and the denominator
\begin{equation}
\begin{split}
D\left(q,y,t\right)=\,&\left(1-t^2\right) \left(1-y^2\right) \left(1-t y\right) \left(1-q^2\right) \left(1-q^3\right) \left(1-q t^2\right) \left(1-q y^2\right) \\
&\left(1-q^2t^2\right) \left(1-q^2y^2\right)\;.
\end{split}
\end{equation}
We have expanded $N$ and $D$ of $\mathfrak{H}(q,y,t)$ to a form in which the leading non-trivial term in $N$ is positive. 
In this form, we observe that the numerator is \textit{anti-palindromic}, $N(q,y,t)=-q^6y^4t^4\,N(q^{-1},y^{-1},t^{-1})$.

Very important information is contained in the multi-graded PL, which is defined as\footnote{%
The original reference for this function seems to be \cite{1994dg.ga.....8003G} and it has also been discussed in \cite{Labastida:2001ts}.
It was introduced in a particle physics context in \cite{Benvenuti:2006qr,Feng:2007ur}.}
\begin{equation}
 \mathrm{PL}\left[\mathfrak{H}\left(q,y,t\right)\right]:=\sum\limits_{k=1}^{\infty}\frac{\mu(k)\, \ln \mathfrak{H}\left(q^k,y^k,t^k\right)}{k}\;,
\end{equation}
where $\mu(k)$ is the M\"obius function.
Expanding the PL around zero for all variables we find 
\begin{equation}\label{eq:PLgraded}
\begin{split}
 \mathrm{PL}\left[\mathfrak{H}\left(q,y,t\right)\right]=\,
 &t^2+t y+y^2+q^2+q t^2+q t y+q y^2+q^3+q t^2 y+q^2 t^2+q t y^2+q^2 t y \\
 &+q^2 y^2+q^2 t^2 y+q^2 t y^2 +q^3 t^3+q^3 t^2 y+q^3 t y^2+q^3 y^3-q^2 t^2 y^2 \\
 &-q^2 t^3 y^2-q^2 t^2 y^3-q^3 t^2 y^2-q^2 t^4 y^2-q^3 t^4 y-q^2 t^3 y^3-3 q^3 t^3 y^2 \\
 & -q^2 t^2 y^4-3 q^3 t^2 y^3-q^4 t^2 y^2-q^3 t y^4-\mathcal{O}\left(\left[tyq\right]^9\right)\;. \raisetag{20pt}
\end{split} 
\end{equation}
The usual, ungraded, HS can directly be obtained from the multi-graded version by equating all arguments\footnote{%
The HS of the 2HDM scalar sector has only very recently appeared in the literature for the first time, see Eq.\ (A.5) of \cite{Bednyakov:2018cmx}. We do not include the 
linear invariants here, which explains the slight difference to \cite{Bednyakov:2018cmx}. Otherwise our results are in full agreement.}
\begin{equation}\label{eq:ungradedHS}
\mathfrak{h}(z)\equiv\mathfrak{H}(z,z,z)= \frac{1+z^3+4\,z^4+2\,z^5+4\,z^6+z^7+z^{10}}{\left(1-z^2\right)^4\left(1-z^3\right)^3\left(1-z^4\right)}\;.
\end{equation}
As expected for a reductive group like \SU2, we can find a form of $\mathfrak{h}(z)$ in which all numerator coefficients are positive.
In this form the numerator of $\mathfrak{h}(z)$ is palindromic, i.e.\ $N(z)=z^{10}N(z^{-1})$.
Finally, the ungraded PL is given by
\begin{equation}\label{eq:PL}
\mathrm{PL}\left[\mathfrak{h}(z)\right]= 4\,z^2+4\,z^3+5\,z^4+2\,z^5+3\,z^6-3\,z^7-\mathcal{O}\left(z^8\right)\;.
\end{equation}
Now, let us point out the relevant information content of these functions for this study:
\begin{itemize}
 \item The denominator of the HS in Eq.\ \eqref{eq:ungradedHS} informs us about the 
 smallest complete set of algebraically independent invariants. 
 We read off that there are four algebraically independent invariants of order 2, three of order 3 and one of order 4.
 \item The leading positive terms of the multi-graded PL in Eq.\ \eqref{eq:PLgraded} correspond to the number and structure of invariants 
 in the generating set of the ring. 
 To be clear, the leading $19$ terms correspond to \textit{all} invariants which are needed to express \textit{any} other invariant as a polynomial of them.
 For example, the leading $t^2$ term tells us that there will be one generating invariant originating from the tensor product $Z_{\rep{3}}\otimes Z_{\rep{3}}$.
 The second term, $ty$, tells us that there will be one invariant from the tensor product $Z_{\rep{3}}\otimes Y_{\rep{3}}$, \textit{etc}.\ .
 \item The leading negative terms of the multi-graded PL in Eq.\ \eqref{eq:PLgraded} cut-off the set of generating invariants. 
 Furthermore, these terms tell us the structure and number of relations between the invariants. For example, the leading 
 negative term, $-q^2y^2t^2$, of the total order $6$ indicates that there is one relation between invariants of that structure.
 The term $-3q^3t^3y^2$ tells us that there are three independent relations between invariants of that structure, \textit{etc}.\ .
\end{itemize}
There is more useful information encoded in these functions but this is the most relevant for the sake of this study.

\subsection{Explicit construction of the invariants}
\label{sec:construction}
\subsubsection{Needed projection operators}

After having found the number and internal structure of the non-linear invariants, we proceed with their explicit construction.
Due to their symmetry properties, we can directly represent the building blocks as symmetrized boxes
\ytableausetup{smalltableaux}
\begin{equation}
 Y_{\rep{3}}~\equiv~\ydiagram[*(darkgreen)]{2}\;,\quad Z_{\rep{3}}~\equiv~\ydiagram[*(red)]{2}\;,\quad Z_{\rep{5}}~\equiv~\ydiagram[*(lightgray)]{4}\;.
\end{equation}
Here we have introduced a color coding to tell apart boxes of the different building blocks.
Arbitrary invariants of higher order are now obtained by taking tensor products of the building blocks 
and projecting out the invariants. It turns out that all required projection operators are very simple and always of the form
\ytableausetup{ boxsize=1.5em}
\begin{equation}\label{eq:GenProjector}
\scalebox{0.7}{\begin{ytableau}
1 & 2 & \none[\cdots] &  n \\
\scriptstyle n+1 & \scriptstyle n+2 &  \none[\cdots] & 2n 
\end{ytableau}}
~\Longrightarrow~
P_{_{
\scalebox{0.4}{
\begin{ytableau}
1 & 2 & \none[\cdots] &  n \\
\scriptstyle n+1 & \scriptstyle n+2 &  \none[\cdots] & 2n 
\end{ytableau}
}}}
~=~
\birdtrack{35ex}{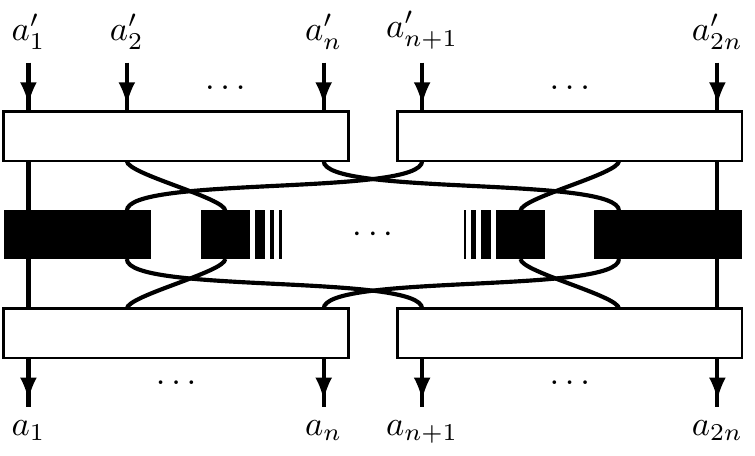}\;.
\end{equation}\ytableausetup{smalltableaux}%
This follows from two facts: 
(i) We are only interested in invariants, and the operators that project onto invariants are precisely the ones which arise from Young tableaux 
of the ``complete chocolate bar'' shape 
(i.e.\ \,\raisebox{2pt}{\scalebox{0.7}{$\begin{ytableau} \phantom{1} & & \none[\scriptstyle\cdots] &  \\ & & \none[\scriptstyle\cdots] &  \end{ytableau}$}}\,).
(ii) Other Young tableaux of the same shape but with a different assignment of indices do exist (e.g.\ \,\raisebox{2pt}{\scalebox{0.7}{$\begin{ytableau}1 & 2 & 3 & 5\\4 & 6 & 7 & 8\end{ytableau}$}}\, instead of
\,\raisebox{2pt}{\scalebox{0.7}{$\begin{ytableau}1 & 2 & 3 & 4\\5 & 6 & 7 & 8\end{ytableau}$}}\,), but their corresponding projection operators 
in our case will not give rise to invariants which are independent of the ones obtained via the projection operators in \eqref{eq:GenProjector}. 

Note that (ii) is \textit{not} true in general, c.f.\ the two independent singlets
extracted from $Z$ via \,\raisebox{2pt}{\scalebox{0.7}{$\begin{ytableau}1 & 2 \\ 3 & 4\end{ytableau}$}}\, and \,\raisebox{2pt}{\scalebox{0.7}{$\begin{ytableau}1 & 3 \\ 2 & 4\end{ytableau}$}}\, above.
Nonetheless, for tensor products of only $Y_{\rep{3}}$, $Z_{\rep{3}}$, and $Z_{\rep{5}}$, the statement (ii) holds and we have confirmed this explicitly. 
This can also be read off directly from the PL \eqref{eq:PLgraded}, where we find that no single invariant structure of the generating set appears with
multiplicity higher than one -- implying that each single tensor product $Z_{\rep{5}}^{\otimes a}\otimes Y_{\rep{3}}^{\otimes b}\otimes Z_{\rep{3}}^{\otimes c}$ 
can host at most one independent invariant.

\subsubsection{Algebraically independent invariants}
\label{sec:primary}
We are now equipped to construct invariants of arbitrary order simply by projection.
Let us introduce the following naming scheme for higher-order invariants: \\
\begin{equation*}
 \bbox{\Inv{a,b,c}
 ~\mathrel{\widehat=}~\text{Invariant containing powers $q^{a}$, $y^{b}$, and $t^{c}$, that is $Z_{\rep{5}}^{\otimes a}\otimes Y_{\rep{3}}^{\otimes b}\otimes Z_{\rep{3}}^{\otimes c}$}.}
\end{equation*}\linebreak
For completeness, we recall the three linear invariants, which are given by
\begin{equation}
 Y_{\rep{1}}~=~
 \begin{ytableau}
*(darkgreen) 1 \\
*(darkgreen) 2
\end{ytableau}\;, \quad
  Z_{\rep[_{(1)}]{1}}~=~
 \begin{ytableau}
*(lightgray) 1 & *(lightgray) 2 \\
*(lightgray) 3 & *(lightgray) 4
\end{ytableau}\;, \quad \text{and} \quad 
Z_{\rep[_{(2)}]{1}}~=~
 \begin{ytableau}
*(lightgray) 1 & *(lightgray) 3 \\
*(lightgray) 2 & *(lightgray) 4
\end{ytableau}\;.
\end{equation}
We proceed with the construction of a set of non-linear independent invariants.
The order of the sought invariants corresponds to the denominator factors of the HS \eqref{eq:ungradedHS},
and their tensor product structure can be read off from the graded PL \eqref{eq:PLgraded}.
The resulting invariants are
\begin{align}\notag
\Inv{2,0,0}:=\ydiagram[*(lightgray)]{4,4}\;,\quad
\Inv{0,2,0}:=\ydiagram[*(darkgreen)]{2,2}\;,\quad
\Inv{0,1,1}:=\ydiagram[*(darkgreen)]{2}*[*(red)]{2,2}\;,\quad
\Inv{0,0,2}:=\ydiagram[*(red)]{2,2}\;,
\end{align}\vspace{-0.8cm}
\begin{align}\notag\label{eq:primaries}
\Inv{3,0,0}&:=\ydiagram[*(lightgray)]{6,6}\;,&
\Inv{1,2,0}&:=\ydiagram[*(lightgray)]{4}*[*(darkgreen)]{4,4}\;,&    \Inv{1,0,2}&:=\ydiagram[*(lightgray)]{4}*[*(red)]{4,4}\;, \quad \text{and}&
\end{align}\vspace{-0.8cm}
\begin{align}
 \Inv{2,1,1}:=\ydiagram[*(lightgray)]{6,2}*[*(red)]{6,2+2}*[*(darkgreen)]{6,4+2}\;.
\end{align}
Here and in the following we suppress indices in the Young tableaux whenever they are meant to be assigned in the trivial way (incremental increase by one from left to right in each line, cf.\ \eqref{eq:GenProjector}).
The corresponding projection operators always have the form stated in \eqref{eq:GenProjector}, and acting with them on the respective tensor product of building blocks
produces the corresponding invariant. Explicit expressions for the invariants obtained in this way are collected in \Appref{App:Invariants} and an 
explicit criterion to check the algebraic independence of arbitrary polynomials is given in \Appref{app:Jacobi}.

Including the linear invariants we have now constructed a total of eleven invariants. These form a maximal set of algebraically independent invariants for the 2HDM.
This number corresponds to the well-known number of $11$ physical parameters of the 2HDM scalar sector. 
As always, the set of algebraically independent invariants is not unique. However, our choice certainly is the simplest
in terms of the order of the individual invariants. 

We see that for the 2HDM scalar sector it is possible to find a maximal set of algebraically independent invariants which are all CP-even.
Just as in the case of the SM \cite{Jenkins:2009dy}, 
this indicates that one can express necessary and sufficient conditions for CP conservation 
solely in terms of CP-even quantities.\footnote{%
In the SM this would be the area of the CKM unitarity triangle reconstructed by the length of its sides.
We thank Jo\~ao P.\ Silva for reminding us of this example.}

\subsubsection{Completing the generating set}
\label{sec:GeneratingSet}

The complete generating set of invariants contains eleven\footnote{%
This number only by accident coincides with the numbers $11$ above, i.e.\ it has nothing to do with the number of physical parameters.}
additional invariants and we collect them here.
After reading off their structure from the graded PL, \Eqref{eq:PLgraded}, the construction proceeds in a straightforward way.
We find
\begin{align}\label{eq:secondaries}\notag
\Inv{1,1,1}&:=\ydiagram[*(lightgray)]{4}*[*(darkgreen)]{4,2}*[*(red)]{4,2+2}\;,&\\[5pt]\notag
\Inv{2,2,0}&:=\ydiagram[*(lightgray)]{6,2}*[*(darkgreen)]{6,6}\;,&             \Inv{2,0,2}&:=\ydiagram[*(lightgray)]{6,2}*[*(red)]{6,6}\;,& \\\notag
\Jnv{1,2,1}&:=\ydiagram[*(lightgray)]{4}*[*(darkgreen)]{5,3}*[*(red)]{5,5}\;,& \Jnv{1,1,2}&:=\ydiagram[*(lightgray)]{4}*[*(red)]{5,3}*[*(darkgreen)]{5,5}\;,& \\[5pt]\notag
\Jnv{2,2,1}&:=\ydiagram[*(lightgray)]{4}*[*(darkgreen)]{7,1}*[*(red)]{7,3}*[*(lightgray)]{7,7}\;,&
\Jnv{2,1,2}&:=\ydiagram[*(lightgray)]{4}*[*(red)]{7,1}*[*(darkgreen)]{7,3}*[*(lightgray)]{7,7}\;,& \\[5pt]\notag
\Jnv{3,3,0}&:=\ydiagram[*(lightgray)]{9,3}*[*(darkgreen)]{9,9}\;,&
\Jnv{3,0,3}&:=\ydiagram[*(lightgray)]{9,3}*[*(red)]{9,9}\;,& \\
\Jnv{3,2,1}&:=\ydiagram[*(lightgray)]{9,3}*[*(darkgreen)]{9,7}*[*(red)]{9,9}\;,& 
\Jnv{3,1,2}&:=\ydiagram[*(lightgray)]{9,3}*[*(red)]{9,7}*[*(darkgreen)]{9,9}\;.&
\end{align}
Invariants that contain an odd total number of $Y_{\rep{3}}$ and $Z_{\rep{3}}$ are CP-odd, cf.\ \Secref{sec:CPproperties}.
and we denote them by the letter $\mathcal{J}$ instead of $\mathcal{I}$.
We give explicit expressions for all of these invariants in \Appref{App:Invariants}.
All of our invariants group into permutation representations under a $Y_{\rep{3}}(y)\leftrightarrow Z_{\rep{3}}(t)$ exchange transformation,
which is, of course, what one would expect because $Y_{\rep{3}}$ and $Z_{\rep{3}}$ behave identically under basis changes.

This completes the construction of the generating set of invariants. Note that the invariants constructed in this section 
are not algebraically independent of the invariants in \Secref{sec:primary} above. Therefore, each 
of these invariants fulfills a polynomial relation with the other invariants and we will now proceed to construct
some of these relations explicitly.

\section{Systematic construction of Syzygies}
\label{sec:syzygies}

By definition of the generating set, it must be possible to express all higher-order invariants as polynomials 
in the invariants listed above. This requires relations between the higher-order invariants and the invariants of the generating set
which are also called syzygies.

We are not aware of any previously stated generally applicable strategy to systematically construct syzygies. 
However, by explicit computation we find that the following strategy works here:

Any of the leading negative terms in the multi-graded PL, \Eqref{eq:PLgraded}, seems to correspond to a new, independent relation
amongst the invariants. The structure of the corresponding term corresponds to the structure of the relation(s), and the coefficient 
of the term gives the number of independent relations of this structure.
Simply making an ansatz of to-be related invariants in terms of suitable power products of lower order invariants, one just has to solve a linear system in order
to obtain the desired syzygie(s). 

For example, the leading negative term in \eqref{eq:PLgraded} is $-q^2y^2t^2$.
All possible power products of invariants that match this structure are
\begin{align}\notag
 &\Inv{1,1,1}^2\;,& &\Inv{2,1,1}\,\Inv{0,1,1}\;,& &\Inv{2,2,0}\,\Inv{0,0,2}\;,& &\Inv{2,0,2}\,\Inv{0,2,0}\;,& \\
 &\Inv{1,2,0}\,\Inv{1,0,2}\;,& &\Inv{2,0,0}\,\Inv{0,2,0}\,\Inv{0,0,2}\;,& &\Inv{2,0,0}\,\Inv{0,1,1}^2\;.&
\end{align}
A simple linear ansatz then reveals the first syzygy:\footnote{%
Comparing this to the corresponding relation in the trace basis, e.g.\ Eq.\ (A.4) in \cite{Bednyakov:2018cmx},
gives a feeling for the simplification arising from the use of orthogonal projectors.
}
\begin{equation}\label{eq:Syz6}
\begin{split}
 3\,\Inv{1,1,1}^2~=&~2\,\Inv{2,1,1}\,\Inv{0,1,1} - \Inv{2,2,0}\,\Inv{0,0,2} - \Inv{2,0,2}\,\Inv{0,2,0} \\ &+ 3\,\Inv{1,2,0}\,\Inv{1,0,2}+\Inv{2,0,0}\,\Inv{0,2,0}\,\Inv{0,0,2}-\Inv{2,0,0}\,\Inv{0,1,1}^2\;.
\end{split}
\end{equation}
In principle it should be possible to find all syzygies by this strategy and we have explicitly checked that this works for invariants up to a total order of $14$,
see \Tabref{tab:Syzygies}.

There is a caveat: Observe the differences between the (expanded) HS coefficients and the PL coefficients of the same term, for example 
\begin{align}\notag
 \mathfrak{H}(q,y,t)~&=~\dots + 6\,q^2 t^2 y^2 + \dots + 10\,q^4 t^2 y^2 + \dots\;, \\ \label{eq:HvsPL}
 \mathrm{PL}\left[\mathfrak{H}\left(q,y,t\right)\right]~&=~ \dots - 1\,q^2 t^2 y^2 + \dots - 1\,\,\,q^4 t^2 y^2 + \dots \;.
\end{align}
For the above example -- $q^2 t^2 y^2$ -- the difference between the HS and PL coefficients \mbox{$6-(-1)=7$} matches the number $7$ of possible power products of the generating invariants.
In some cases, however, the number of possible power products exceeds this difference. 
For example, this happens for the invariants of order $q^4 t^2 y^2$, for which according to the PL coefficient there is one non-trivial relation. 
The difference in the coefficients between HS and PL here is $11$ -- but we find that there are $12$ possible power products.
This is indicative of an additional relation of the structure $q^4 t^2 y^2$, in addition to  the one counted by the ``$-1$'' in the PL, \Eqref{eq:HvsPL}. 
However, this additional relation turns out not to be independent of the other relations.
Rather, it is the ``old'' $q^2 t^2 y^2$ relation \Eqref{eq:Syz6} multiplied by the invariant $q^2$, thus producing a relation of the order $q^4 t^2 y^2$. 
While this happening does not give an obstacle for the explicit construction of independent syzygies, one should keep it in mind for explicit computations.

The structure of all relations that we have explicitly constructed and checked in this way are shown in \Tabref{tab:Syzygies} in 
\Appref{app:syzygies}.
Altogether this gives some evidence to the suspicion that one can simply read off the total number of independent relations 
and their structure from the negative terms of the PL. 

\section{Necessary and sufficient conditions for explicit CPV}
\label{sec:ConditionsforCPC}
Let us now make use of our newly gained knowledge about the 2HDMs invariants.
In a formidable explicit computation \cite{Gunion:2005ja} it has been shown that an equivalent condition to explicit CP 
conservation in the 2HDM scalar sector is the vanishing of the four specific CP-odd basis invariants
\begin{align}\label{eq:CPinvariants}\notag
 I_{2Y2Z}~&\sim~\Jnv{1,2,1}\;,&  I_{Y3Z}~&\sim~\Jnv{1,1,2}\;,& \\
 I_{3Y3Z}~&\sim~\Jnv{3,3,0}\;,& I_{6Z}~&\sim~\Jnv{3,0,3}\;.&   
\end{align}
We have also stated here the corresponding invariants in the notation of this paper.
``$\sim$'' here means that the non-trivial parts of the invariants coincide while their exact expressions 
may differ by a global numerical prefactor and an admixture of lower-order invariants of the correct structure.
The exact relations are given in \Appref{app:GHInvariants}.

We will now give a very simple proof of the necessary and sufficient conditions for explicit CPV by making use of the interrelation of the invariants.

The first step of this proof is the insight that instead of a potentially infinite number of CP-odd invariants
we only have to deal with the CP-odd invariants in the generating set of the ring.
This is clear from the fact that any other invariant can be expressed as a polynomial in these.

From \Eqref{eq:secondaries} we then find that there are eight CP-odd invariants in the generating set of the ring.
Thus, in order to prove Gunion and Haber right, there should be at least four independent relations amongst these invariants.
In fact, we find that there are many more relations among the CP-odd invariants, cf.\ \Tabref{tab:Syzygies}.

The first two relations are arising at a total order $7$ and they read
\begin{equation}\label{eq:CPoddSyz7}
 \Jnv{2,2,1} \Inv{0,0,2} + \Jnv{2,1,2} \Inv{0,1,1} - \Jnv{1,2,1} \Inv{1,0,2} + \Jnv{1,1,2} \Inv {1,1,1}~=~0\;, \quad\text{and}\quad y\leftrightarrow t\;.
\end{equation}
Note how this is a fully ``CP-odd relation'' which is actually sensitive to the signs of the CP-odd invariants and not only to their magnitude.
This relation will not be used in the proof but we have stated it for completeness. 
Furthermore, we find two CP-odd relations of total order $8$ that read
\begin{equation}\label{eq:CPoddSyz8}
3\,\Jnv{2,2,1} \Inv{1,2,0} - \Jnv{3,2,1} \Inv{0,2,0} + 3\,\Jnv{3,3,0} \Inv{0,1,1} + \Jnv{1,2,1} \Inv{2,2,0}~=~0\;, \quad\text{and}\quad y\leftrightarrow t\;.
\end{equation}
Finally, we also state the ``CP-even'' relations of the squared quintic CP-odd invariants
\begin{equation}\label{eq:CPevenSyz10}
3\,\Jnv{2,2,1}^2 + 3\,\Jnv{3,3,0} \Jnv{1,1,2} - \Jnv{3,2,1} \Jnv{1,2,1} -\Jnv{1,2,1}^2 \Inv{2,0,0}~=~0\;, \quad\text{and}\quad y\leftrightarrow t\;.
\end{equation}
Remarkably, the squared quintics can be expressed almost entirely in terms of products of the other CP-odd invariants.

Many more relations exist and have been derived, see \Tabref{tab:Syzygies}, but they are not needed for the proof and so we do not state them here explicitly.

Let us now show that the vanishing of the invariants in \eqref{eq:CPinvariants}, that is  
\begin{equation}\label{eq:sufficient}
\Jnv{1,2,1}~=~\Jnv{1,1,2}~=~\Jnv{3,3,0}~=~\Jnv{3,0,3}~=~0, 
\end{equation}
is sufficient to conclude that all other CP-odd invariants are vanishing as well. 
This is readily confirmed by inspection of Eqs.~\eqref{eq:CPoddSyz8} and \eqref{eq:CPevenSyz10}.
Using \eqref{eq:sufficient} it follows from \eqref{eq:CPevenSyz10} that also $\Jnv{2,2,1}=\Jnv{2,1,2}=0$.
Using this together with the condition \eqref{eq:sufficient} in \eqref{eq:CPoddSyz8} one finds that $\Jnv{3,2,1}=\Jnv{3,1,2}=0$, but only under the assumption that $\Inv{0,2,0}\neq0\neq\Inv{0,0,2}$.
However, if $\Inv{0,2,0}=\Inv{0,0,2}=0$ \emph{were to hold}, this would itself imply that $\Jnv{3,2,1}=\Jnv{3,1,2}=0$ to begin with, which is easy to show from
the explicit form of the invariants given in \Appref{App:Invariants}. This completes the proof.

The in total $6$ novel relations \eqref{eq:CPoddSyz7}, \eqref{eq:CPoddSyz8} and \eqref{eq:CPevenSyz10} are perhaps the main results of this paper. 
However, many more relations of this type exist, and so we are convinced that the main use of this work is the way of how we got there.

\section{Towards a Hironaka decomposition}
\label{sec:Hiroanaka}
Finally, we wish to comment on the possibility of representing the ring of invariants of the 2HDM in an even simpler way.
Due to the fact that the \SU2 group of basis changes is reductive, it follows that the ring of invariants
obeys the Cohen-Macaulay (CM) property implying the existence of a so-called Hironaka decomposition \cite{Hochster:1974a, Hochster:1974b} 
(c.f.\ also \cite[Sec.~2.3]{Sturmfels:2008}, \cite[Sec.~5.4.1]{Henning:2017fpj},\cite[App.~A]{Bourget:2018ond} or \cite[Sec.~3]{2008arXiv0812.3082T}).

Let us explain what that means. Writing the HS of \Eqref{eq:ungradedHS} as%
\begin{equation}\label{eq:GeneralHS}
 \mathfrak{h}(z)=\frac{\sum\limits_{i=1}^{s}z^{s_i}}{\prod\limits_{j=1}^{p}\left(1-z^{p_j}\right)}\;,
\end{equation}
defines the numbers $p$, $p_j$, $s$, and $s_i$.\footnote{%
For the case of \Eqref{eq:ungradedHS} one finds $p=8$, $s=14$ as well as the corresponding orders of the invariants $p_j=\{2,2,2,2,3,3,3,4\}$, $s_i=\{0,3,4,4,4,4,5,5,6,6,6,6,7,10\}$.}
The CM property then warrants that the ring can be generated in terms of a number of $p$ \textit{primary} invariants of the orders $p_j$, commonly denoted as $\theta_j$,
together with a number of $s$ \textit{secondary} invariants of the orders $s_i$, commonly denoted as $\eta_i$.
Together, the primary and secondary invariants form a Hironaka decomposition of the ring, implying that \textit{every} invariant $\mathcal{I}$ can be written as%
\begin{equation}\label{eq:hironaka}
  \mathcal{I}=\sum\limits_{i=1}^{s} \eta_i\,\mathbb{C}\left[\theta_1,\dots,\theta_p\right]\;.
\end{equation}
Here $\mathbb{C}\left[\theta_1,\dots,\theta_p\right]$ denotes arbitrary polynomials in the primary invariants with potentially complex coefficients and
$\eta_i$ are the secondary invariants.
The astonishing feature of this decomposition is that the secondary invariants only enter linearly. 

Using the Hironaka decomposition can simplify the analysis of the ring and the related physical discussion.
For example, imagine the situation that all primary invariants of a ring are CP-even, while CP-odd invariants only arise as secondary invariants.
In order to find necessary and sufficient conditions for CP conservation, one then would only have to investigate the linear span of the CP-odd
secondary invariants (with coefficients in the primary invariants). 

We stress that the power product invariants stated above in \Secref{sec:primary} and \Secref{sec:GeneratingSet} 
do not directly correspond to the primary and secondary invariants of the Hironaka decomposition. 
Thus, given the generating set of a ring, the main task in obtaining a Hironaka decomposition is the identification of primary and secondary invariants.
This can be complicated by the fact that some of the primary and secondary invariants may only arise as combinations of invariants in the generating set.
Whether this happens is dictated by the interrelations of invariants.
In particular, any product of secondary invariants must \textit{either} decompose in the form \eqref{eq:hironaka}
\textit{or} be a secondary invariant itself. 
The latter situation explains the appearance of secondary invariants of degree larger than any of the invariants in the generating set.
These higher-order, so-called ``reducible'' secondary invariants can always be written as power 
products of a set of ``irreducible'' secondary invariants \cite{2007math......1270K}.

We now comment on the construction of a Hironaka decomposition of the 2HDM ring.
The primary invariants of order $2$ and $3$ are simply those listed in \eqref{eq:primaries}.
The remaining cubic invariant $\Inv{1,1,1}$ fulfills the relation \eqref{eq:Syz6}, rendering it a bona-fide secondary invariant.
The main obstacle, thus, is the identification of the lone degree $4$ primary invariant.
It is clear that it must be a combination of the quartic invariants given in \eqref{eq:primaries} and \eqref{eq:secondaries}.
Since there are no syzygies of the structure $(q^2y^2)^2$, or $(q^2t^2)^2$, and since there are no secondary invariants of order $8$,
both of the invariants $\Inv{2,2,0}$, and $\Inv{2,0,2}$ must be part of the sought primary invariant.
In addition, the syzygy of the structure $(q^2yt)^2$ contains both, a term $\Inv{2,1,1}^2$ as well as a term $\Inv{2,2,0}\times\Inv{2,0,2}$,
thereby indicating that also the invariant $\Inv{2,1,1}$ should appear in the sought quartic primary invariant.
Other combinations of these invariants, together with the remaining quartic invariants $\Inv{1,2,1}$ and $\Inv{1,1,2}$
then should form the four quartic secondary invariants. Together, the sought primary and secondary invariants must obey the Hironaka decomposition \eqref{eq:hironaka}.
Trying to satisfy this requirement with an ansatz, however, leads to a complicated non-linear system of equations which we were unable to solve.
The task of obtaining a Hironaka decomposition of the 2HDM ring, thus, remains to be solved in future work. 

\section{Summary and Discussion}

We have presented a new algorithm to systematically construct basis invariants.
Our method is based on algebraic invariant theory, with the powerful Hilbert series and Plethystic logarithm,
as well as group theory, with Young tableaux and their corresponding hermitian projection operators.

Applied to the 2HDM, we have obtained a maximal set of algebraically independent invariants,
as well as a complete generating set of invariants. We have also outlined a systematic
approach for the construction of syzygies and showed some of them explicitly.
Using the invariants of the generating set and their syzygies we have given a concise proof of the 
necessary and sufficient conditions for explicit CP conservation in the 2HDM.

It should be remarked that the construction of building blocks and invariants in the 2HDM has also been performed by other means.
For example, the building blocks of the present work correspond to the vectors and matrices obtained using the ``bilinear formalism'' 
\cite{Nagel:2004sw,Ivanov:2005hg,Maniatis:2006fs, Ivanov:2006yq, Maniatis:2007vn, Ivanov:2007de,Ferreira:2010hy, Ferreira:2010yh, Maniatis:2014oza, Ivanov:2014doa},
and some of our CP-odd invariants are closely related to the ones obtained via graphical methods in \cite{Gunion:2005ja,Davidson:2005cw}.
There are some advantages of our method as compared to previous approaches. 
First, our strategy does not require specific basis choices at any step. 
Then, by the use of the powerful invariant theory functions we can with certainty identify the
full set of generating invariants, their structure, as well as their interrelations.
This is the first method where the answer to the question ``when to stop constructing invariants'' is
given in a clear and quantitative fashion; it can be calculated in any model from our algorithm 
without the fear of miscounting.
Furthermore, thanks to the use of orthogonal projection operators,
our basis invariants are as short as possible by construction, and their CP properties are very transparent.
Another attractive feature is the direct access to syzygies and their simple form of appearance.

An immediate further application of our method to the 2HDM would be the inclusion of vacuum expectation values
to investigate possibly realistic models and the spontaneous violation of CP. 
Our method also includes the possibility of identifying how the invariants
behave under the various additional Higgs-flavor symmetries which are all subgroups of the
group of basis changes. As a result, one would be able to identify how the number of independent invariants
is reduced in models which are not the most general, but which have additional internal symmetries.
Using our short invariants and relations, simplifications should also arise in the formulation of the renormalization group running \cite{Bednyakov:2018cmx, Bijnens:2018rqw}.
Another remaining open question in the 2HDM is the identification of combinations of invariants
to obtain a Hironaka decomposition of the ring.

Perhaps more important than our explicit results on the 2HDM is the method itself.
In principle, our strategy generalizes to the fermion sector, three or even $N$-Higgs-doublet models, and also to completely different models.
A possible issue then could become computational power, since the construction of large
Young projection operators easily exhausts memory limits of commercially available computing clusters.
Also the computation of the Hilbert series and Plethystic logarithm straightforwardly extends to more complicated groups, 
but the corresponding integrals become more involved.
Finally, the use of Young tableaux, at a first glance, may seem to limit our explicit construction of invariants to \SU{N} groups. 
However, all that is actually needed for the systematic construction of building blocks and invariants are the hermitian projection operators 
for the various tensor contractions. 
Since birdtracks can be used to construct these for all simple Lie groups \cite{Cvitanovic:2008zz},
we are optimistic that our method for the construction of invariants can also be generalized in this direction.

\subsection*{Acknowledgments}
I want to thank I.\ P.\ Ivanov, C.\ C.\ Nishi, and J.\ P.\ Silva for useful discussions and motivation,
and I.\ P.\ Ivanov and J.\ P.\ Silva for useful comments on the manuscript.
Furthermore, I want to thank the Centre for Theoretical Particle Physics (CFTP) at T\'ecnico Lisboa for hospitality 
at various stages of this work.
This work has been supported by the German
Science Foundation (DFG) within the SFB-Transregio TR33 ``The Dark Universe''.
The author is grateful to the authors of the \LaTeX\ packages $\mathtt{ytableau}$ and $\mathtt{youngtab}$,
and would also like to encourage the development of packages to typeset birdtracks and birdtrack projection operators.

\appendix
\section{Algebraic (in)dependence of invariants}
\label{app:Jacobi}
The Jacobi criterion is an easy way to find the number of algebraically independent polynomials.\footnote{%
Strictly speaking this works only for polynomials over fields of characteristic zero, see e.g.\ \cite{Ehrenborg:1993,Beecken:2011}.}
For a set of polynomials (here, invariants) $\mathcal{I}_i$, depending on a number of variables $x_j$ (here, the components $y_{ab}$ and $z_{abcd}$ of $Y$ and $Z$), the number of algebraically independent invariants is simply given by
the rank of the Jacobian matrix:
\begin{equation}
\text{Number of algebraically independent invariants}~=~\rank \left[ \frac{\partial\,\mathcal{I}_i}{\partial x_j} \right] \;.
\end{equation}
Besides symbolic evaluation one can also use this criterion with all variables put to random numbers for a fast machine evaluation. 

\section{Building blocks in conventional notation}
\label{app:LambdaNotation}
For completeness, we collect here the building blocks constructed in \Secref{sec:buildingblocks}~in the conventional parametrization
of the 2HDM scalar potential in terms of $m$'s and $\lambda$'s.
Following the replacement rules in \cite[Eq.\ (21)]{Gunion:2005ja} we find
(note that their $Z_{a\overline{b}c\overline{d}}\equiv 2\,Z\indices{^{ac}_{bd}}$ in our notation)
\begin{align}\notag
 &y_{11}=m_{11}^2\;,& &y_{12}=-m_{12}^2\;,& &y_{22}=m_{22}^2\;,& \\ \notag
 &z_{1111}=\lambda_1/2\;,& &z_{2222}=\lambda_2/2\;,& &z_{1212}=\lambda_3/2\;,& &z_{1221}=\lambda_4/2\;,& \\
 &z_{1122}=\lambda_5/2\;,& &z_{1112}=\lambda_6/2\;& &z_{1222}=\lambda_7/2\;.&
\end{align}
As usual $\left\{m_{12}^2,\lambda_5,\lambda_6,\lambda_7\right\}\in\mathbb{C}$ while all others are real.
It is then straightforward to translate \Eqref{eq:buildingblocks} into this convention and we find
\begin{subequations}\label{eq:buildingblocksLambdaNotation}
\begin{align}
 Y_{\rep[_{\phantom{{(1)}}}]{1}}&~=~m_{11}^2+m_{22}^2\;,& \\
 Z_{\rep[_{(1)}]{1}}&~=~\frac12\left(\lambda_1+\lambda_2+\lambda_3+\lambda_4\right)\;,& \\
 Z_{\rep[_{(2)}]{1}}&~=~\frac12\left(\lambda_3-\lambda_4\right)\;,& \\
 Y_{\rep{3}}^{ab}&~=~\left(
 \begin{array}{cc}
 -m_{12}^2 & \frac{1}{2} \left(m_{22}^2-m_{11}^2\right) \\
 \frac{1}{2} \left(m_{22}^2-m_{11}^2\right) & (m_{12}^2)^* \\
 \end{array}
 \right)\;,& \\ 
 Z_{\rep{3}}^{ab}&~=~\frac12\left(
 \begin{array}{cc}
 \phantom{\frac12}\,\lambda_6+\lambda_7 & \frac{1}{2} \left(\lambda_2-\lambda_1\right) \\
 \frac{1}{2} \left(\lambda_2-\lambda_1\right) & \,-\left(\lambda_6+\lambda_7\right)^* \\
 \end{array}
 \right)\;,&\\ 
 Z_{\rep{5}}^{abcd}&~=~\left(
\begin{array}{cc}
 \left(
\begin{array}{cc}
 \phantom{-}\xi_1 & \phantom{-}\xi_2 \\
 \phantom{-}\xi_2 & \phantom{-}\xi_3 \\
\end{array}
\right) & \left(
\begin{array}{cc}
 \phantom{-}\xi_2 & \phantom{-}\xi_3  \\
 \phantom{-}\xi_3 & -\xi_2^* \\
\end{array}
\right) \\[0.3cm]
 \left(
\begin{array}{cc}
 \phantom{-}\xi_2 & \phantom{-}\xi_3  \\
 \phantom{-}\xi_3 & -\xi_2^* \\
\end{array}
\right) & \left(
\begin{array}{cc}
 \phantom{-}\xi_3 & -\xi_2^*  \\
 -\xi_2^* & \phantom{-}\xi_1^* \\
\end{array}
\right) \\
\end{array}
\right)
 \;,&
\end{align}
\end{subequations}
with 
\begin{align}
 \xi_1&:=\frac12\lambda_5\;,& 
 \xi_2&:= \frac{1}{4} \left(\lambda_7-\lambda_6\right)\;,& 
 \xi_3&:=\frac{1}{12} \left(\lambda_1-2 \lambda_3-2 \lambda_4+\lambda_2\right)\;.&
\end{align}

\section{Relation to the invariants of Gunion and Haber}
\label{app:GHInvariants}

Here we state the original set of necessary and sufficient CP-odd invariants of Gunion and Haber \cite[Eq.\ (23)-(26)]{Gunion:2005ja} 
in the notation of this paper, 
\begin{align}\label{eq:GHInvariants}
 I_{Y3Z}&=\im\left[Z\indices{^{ai}_{ic}}Z\indices{^{ej}_{jb}}Z\indices{^{bc}_{ed}}Y\indices{^d_a}\right]=-2 \I\,\Jnv{1,1,2}\;,& \\
 I_{2Y2Z}&=\im\left[Y\indices{^{a}_{b}}Y\indices{^{c}_{d}}Z\indices{^{bd}_{af}}Z\indices{^{fi}_{ic}}\right]=-2 \I\,\Jnv{1,2,1}\;,& \\
 I_{6Z}&=\im\left[Z\indices{^{ac}_{bd}}Z\indices{^{b\ell}_{\ell f}}Z\indices{^{dp}_{ph}}Z\indices{^{fj}_{ak}}Z\indices{^{km}_{jn}}Z\indices{^{nh}_{mc}}\right]=-2 \I\,\Jnv{3,0,3}\;,& \\\notag
 I_{3Y3Z}&=\im\left[Z\indices{^{ab}_{cd}}Z\indices{^{cd}_{eg}}Z\indices{^{ef}_{hq}}Y\indices{^{g}_{a}}Y\indices{^{h}_{b}}Y\indices{^{q}_{f}}\right]& \\
 &=2\I\,\Jnv{3,3,0}+2\I\,\Jnv{1,2,1}\,\Inv{0,1,1}+\I\,Y_{\rep{1}}^2\,\Jnv{1,1,2}\;.&
 \end{align}
The first three invariants only differ by a global numerical factor, while the fourth invariant of \cite{Gunion:2005ja}
actually contains admixtures of lower lying invariants.
Note also that our invariants are pure complex numbers by construction. 
That is, they directly correspond to the imaginary, CP-odd part of a given contraction.

\section{Explicit statement of the Invariants}
\label{App:Invariants} 

\subsection{Algebraically independent invariants}
In this appendix we state explicitly our choice of a maximal set of algebraically independent basis invariants of the 2HDM.
Note that these invariants are not written in any specific basis, i.e.\ they hold for all bases.

To ease the notation we use redefined parameters (conserving the total number of degrees of freedom, of course) 
which reflect the now known assignment of the Lagrangian parameters in $Y$ and $Z$ into the building blocks.
Furthermore, it is convenient to split parameters into their real and imaginary parts.

In our new conventions the building blocks of \Eqref{eq:buildingblocks} read
\begin{subequations}\label{eq:buildingblocks2}
\begin{align}
 Y_{\rep[_{\phantom{{(1)}}}]{1}}:&~y_{\mathrm{s}}\;,& \\
 Z_{\rep[_{(1)}]{1}}:&~s_1\;,& \\
 Z_{\rep[_{(2)}]{1}}:&~s_2\;,& \\
 Y_{\rep{3}}^{ab}:&~\left(
 \begin{array}{cc}
 \I\,\yi+\yr & y \\
 y & \I\,\yi-\yr \\
 \end{array}
 \right)\;,& \\  
 Z_{\rep{3}}^{ab}:&~\left(
 \begin{array}{cc}
 \I\,\ti+\tr & t \\
 t & \I\,\ti-\tr \\
 \end{array}
 \right)\;,&\\ 
 Z_{\rep{5}}^{abcd}:&~
 \left(
\begin{array}{cc}
 \left(
\begin{array}{cc}
 \I\,\qi1+\qr1 & \I\,\qi2+\qr2 \\
 \I\,\qi2+\qr2 & q_3 \\
\end{array}
\right) & \left(
\begin{array}{cc}
 \I\,\qi2+\qr2 & q_3 \\
 q_3 & \I\,\qi2-\qr2 \\
\end{array}
\right) \\
 \left(
\begin{array}{cc}
 \I\,\qi2+\qr2 & q_3 \\
 q_3 & \I\,\qi2-\qr2 \\
\end{array}
\right) & \left(
\begin{array}{cc}
 q_3 & \I\,\qi2-\qr2 \\
 \I\,\qi2-\qr2 & \qr1-\I\,\qi1 \\
\end{array}
\right) \\
\end{array}
\right)
 \;,&
\end{align}
\end{subequations}
where now all fourteen parameters $\left\{y_{\mathrm{s}},s_1,s_2,y,\yr,\yi,t,\tr,\ti,\qr1,\qi1,\qr2,\qi2,q_3\right\}$ are real.
Their expressions in terms of the original parameters can easily be obtained by comparison with \Eqref{eq:buildingblocks} or \Eqref{eq:buildingblocksLambdaNotation}.

The global prefactors of invariants are irrelevant for our purpose.
However, the convention for the prefactors does, of course, matter in stating syzygies such as the ones in \Eqref{eq:Syz6}, 
\eqref{eq:CPoddSyz7}, \eqref{eq:CPoddSyz8}, or \eqref{eq:CPevenSyz10}.
As a convention, we adjust the global prefactor of each invariant such as to render all internal coefficients integer. 
This choice also makes the relative coefficients in the syzygies very simple.

Explicit expressions for the algebraically independent invariants of \Secref{sec:primary} are 
\small
\begin{subequations}
\begin{align}\notag
Y_{\rep{1}}\;\;\;~&:=~\,
\begin{ytableau}
*(darkgreen) 1 \\
*(darkgreen) 2
\end{ytableau}~= 
\left(y_{11}+y_{22}\right)~\equiv~ y_s,&\\ \notag
Z_{\rep[_{(1)}]{1}}~&:=~\,
\begin{ytableau}
*(lightgray) 1 & *(lightgray) 2 \\
*(lightgray) 3 & *(lightgray) 4
\end{ytableau}~= 
\left(z_{1111}+z_{1212}+z_{1221}+z_{2222}\right)~\equiv~ s_1,&\\
Z_{\rep[_{(2)}]{1}}~&:=~\,
\begin{ytableau}
*(lightgray) 1 & *(lightgray) 3 \\
*(lightgray) 2 & *(lightgray) 4
\end{ytableau}~= \left(z_{1212}-z_{1221}\right)~\equiv~s_2,&
\end{align}%
\begin{align}\notag
\Inv{0,2,0}~&:=~
\ydiagram[*(darkgreen)]{2,2}~=
  \left(y^2+\yi^2+\yr^2\right),&\\ \notag
\Inv{0,1,1}~&:=~
  \ydiagram[*(red)]{2}*[*(darkgreen)]{2,2}~=
  \left(t y+t_{\I} \yi+\tr \yr\right),&\\ \notag
\Inv{0,0,2}~&:=~
  \ydiagram[*(red)]{2,2}~=
  \left(t^2+t_{\I}^2+\tr^2\right), & \\
\Inv{2,0,0}~&:=~
\ydiagram[*(lightgray)]{4,4}~=
  \left(3 q_3^2+\qi1^2+4 \qi2^2+\qr1^2+4 \qr2^2\right),&
\end{align}%
\begin{align}\notag
\Inv{1,2,0}~&:=~
\ydiagram[*(lightgray)]{4}*[*(darkgreen)]{4,4}~=
  \left[q_3 \left(2 y^2-\yi^2-\yr^2\right)+4 y \left(\qi2 \yi 
    +\qr2 \yr\right) + 2 \qi1 \yi \yr+\qr1 \left(\yr^2-\yi^2\right)\right],& \\ \notag
\Inv{1,0,2}~&:=~
\ydiagram[*(lightgray)]{4}*[*(red)]{4,4}~=
  \left[q_3 \left(2 t^2- t_{\I}^2- \tr^2\right)+4 t \left(\qi2 t_{\I} 
    +\qr2 \tr\right)+2 \qi1 t_{\I} \tr+\qr1\left(\tr^2 -t_{\I}^2\right)\right],&\\ 
\Inv{3,0,0}~&:=~
\ydiagram[*(lightgray)]{6,6}~=
    \left(-q_3 \left(\qi1^2-2 \qi2^2+\qr1^2-2 \qr2^2\right)+
    q_3^3-2 \qi2^2 \qr1+4 \qi1 \qi2 \qr2+2 \qr1 \qr2^2\right),&
\end{align}%
\vspace{-0.5cm}
\begin{equation}
\begin{split}
\Inv{2,1,1}~:=~
\ydiagram[*(lightgray)]{6,2}*[*(red)]{6,2+2}*[*(darkgreen)]{6,4+2}~=
&\left\{ 6 q_3 \left[-\qi2 (t \yi+t_{\I} y)+\qi1 (t_{\I} \yr+\tr \yi)+\qr1 (\tr \yr-t_{\I} \yi)\right] \right. \\ 
& -6 \qr2 \left[ q_3 (t \yr+\tr y)+\qi1 (t \yi+t_{\I} y)+2 \qi2 (t_{\I} \yr+\tr \yi)+\qr1 (t \yr+\tr y)\right] \\
& +3 q_3^2 (-2 t y+t_{\I} \yi+\tr \yr)+6 \qi2 \qr1 (t \yi+t_{\I} y) \\
& +\qi1^2 (2 t y-t_{\I} \yi-\tr \yr)-4 \qi2^2 (t y+t_{\I} \yi-2 \tr \yr)-6 \qi1 \qi2 (t \yr+\tr y) \\
&\left. +\,\qr1^2 (2 t y-t_{\I} \yi-\tr \yr)-4 \qr2^2 (t y-2 t_{\I} \yi+\tr \yr)\right\}.&
\raisetag{15pt}
\end{split}
\end{equation}
\end{subequations}
\normalsize

\subsection{Invariants to complete the generating set}
Here we give explicit expressions for the invariants of \Secref{sec:GeneratingSet}.
The statement in text-form here is provided for completeness. For practical applications we 
also provide the invariants in an auxiliary \texttt{Mathematica} notebook.

\scriptsize
\begin{equation}
\begin{split}
\ydiagram[*(lightgray)]{4}*[*(red)]{4,2}*[*(darkgreen)]{4,2+2}~=~
  &
  \left[q_3\left(2 t y-t_{\I} \yi-\tr \yr\right)
  + 2\qi2 \left(t \yi+ t_{\I} y\right) \right. \\  
  &\left. + \qi1 \left(t_{\I} \yr+ \tr \yi\right)
  + 2\qr2 \left(t \yr+ \tr y\right)
  + \qr1 \left(\tr \yr-t_{\I} \yi\right) \right],&
\end{split}
\end{equation}\vspace{-0.3cm}
 \begin{equation}
\begin{split}
\ydiagram[*(lightgray)]{4}*[*(darkgreen)]{5,3}*[*(red)]{5,5}~=~\I 
 &\left[3 q_3 y (\tr \yi-\ti \yr)\right. \\[-5pt]
 &+2 \qi2 \left(t y \yr-\ti \yi \yr-\tr y^2+\tr \yi^2\right)
  +\qi1 \left(t \left(\yr^2-\yi^2\right)+\ti y \yi-\tr y \yr\right) \\
 &+\qr1 (-2 t \yi \yr+\ti y \yr+\tr y \yi) \\
 &\left.+\,2 \qr2 (-t y \yi+\ti (y-\yr) (y+\yr)+\tr \yi \yr)\right], \\
\end{split}
\end{equation}\vspace{-0.3cm}
\begin{equation}
\begin{split}
\ydiagram[*(lightgray)]{4}*[*(red)]{5,3}*[*(darkgreen)]{5,5}~=~-\I
 &\left[3 q_3 t (\tr \yi-\ti \yr) \right.\\[-5pt]
 &+2 \qi2 \left(t^2 \yr-t \tr y+\ti (\tr \yi-\ti \yr)\right)
  +\qi1 \left(-t \ti \yi+\tr (t \yr-\tr y)+\ti^2 y\right) \\
 &+2 \qr2 \left(-t^2\yi+t \ti y+\tr (\tr \yi-\ti \yr)\right) \\
 &\left.+\,\qr1 (-t \ti \yr-t \tr \yi+2 \ti \tr y)\right],\\
\end{split}
\end{equation}\vspace{-0.3cm}
\begin{align}
\begin{autobreak}\MoveEqLeft
\ydiagram[*(lightgray)]{6,2}*[*(darkgreen)]{6,6}~=~ 
-24 \qi2 \qr2 \yi \yr
-12 \left(q_3 \qi2 y \yi-q_3 \qi1 \yi \yr+q_3 \qr2 y \yr-\qi2 \qr1 y \yi+\qi1 \qr2 y \yi+\qi1 \qi2 y \yr+\qr1 \qr2 y \yr\right)
+8 \left(\qi2^2 \yr^2+\qr2^2 \yi^2\right)
-6 \left(q_3 \qr1 \yi^2-q_3 \qr1 \yr^2+q_3^2 y^2\right)
-4 \left(\qi2^2 y^2+\qi2^2 \yi^2+\qr2^2 y^2+\qr2^2 \yr^2\right)
+3 \left(q_3^2 \yi^2+q_3^2 \yr^2\right)
+2 \left(\qi1^2 y^2+\qr1^2 y^2\right)
-\qi1^2 \yi^2-\qi1^2 \yr^2-\qr1^2 \yi^2-\qr1^2 \yr^2
\end{autobreak}\\[0.2cm]
\begin{autobreak}\MoveEqLeft
\ydiagram[*(lightgray)]{6,2}*[*(red)]{6,6}~=~  
-24 \qi2 \qr2 \ti \tr
-12 \left(q_3 \qi2 t \ti-q_3 \qi1 \ti \tr+q_3 \qr2 t \tr-\qi2 \qr1 t \ti+\qi1 \qr2 t \ti+\qi1 \qi2 t \tr+\qr1 \qr2 t \tr\right)
+8 \left(\qi2^2 \tr^2+\qr2^2 \ti^2\right)
-6 \left(q_3 \qr1 \ti^2-q_3 \qr1 \tr^2+q_3^2 t^2\right)
-4 \left(\qi2^2 t^2+\qi2^2 \ti^2+\qr2^2 t^2+\qr2^2 \tr^2\right)
+3 \left(q_3^2 \ti^2+q_3^2 \tr^2\right)
+2 \left(\qi1^2 t^2+\qr1^2 t^2\right)
-\qi1^2 \ti^2-\qi1^2 \tr^2-\qr1^2 \ti^2-\qr1^2 \tr^2
\end{autobreak} \\[0.2cm]
\begin{autobreak}\MoveEqLeft
\ydiagram[*(lightgray)]{4}*[*(darkgreen)]{7,1}*[*(red)]{7,3}*[*(lightgray)]{7,7}~=~ \I \left[ \right. 
4 \left(-q_3 \qr1 t \yi \yr\right.
  +\qi2 \qr2 t \yi^2
  -\qi2 \qr2 t \yr^2
  -\qi2 \qr2 \ti y \yi
  +\qi2 \qr2 \tr y \yr
  -\qi2^2 t \yi \yr
  +\qi2^2 \ti y \yr
  +\qr2^2 t \yi \yr
  \left.-\qr2^2 \tr y \yi\right)
+3 \left(q_3^2 \ti y \yr-q_3^2 \tr y \yi\right)
+2 \left(-q_3 \qi2 t y \yr \right.
  -q_3 \qi1 t \yi^2
  +q_3 \qi1 t \yr^2
  +q_3 \qi1 \ti y \yi
  +q_3 \qi2 \ti \yi \yr
  +q_3 \qi2 \tr y^2
  -q_3 \qi1 \tr y \yr
  -q_3 \qi2 \tr \yi^2
  +q_3 \qr2 t y \yi
  -q_3 \qr2 \ti y^2
  +q_3 \qr1 \ti y \yr
  +q_3 \qr2 \ti \yr^2
  +q_3 \qr1 \tr y \yi
  -q_3 \qr2 \tr \yi \yr
  -\qi1 \qr2 t y \yr
  +\qi2 \qr1 t y \yr
  -\qi2 \qr1 \ti \yi \yr
  +\qi1 \qr2 \ti \yi \yr
  -\qi2 \qr1 \tr y^2
  +\qi1 \qr2 \tr y^2
  -\qi1 \qr2 \tr \yi^2
  +\qi2 \qr1 \tr \yi^2
  +\qi1 \qi2 t y \yi
  -\qi1 \qi2 \ti y^2
  +\qi1 \qi2 \ti \yr^2
  -\qi1 \qi2 \tr \yi \yr
  +\qr1 \qr2 t y \yi
  -\qr1 \qr2 \ti y^2
  +\qr1 \qr2 \ti \yr^2
  \left.-\qr1 \qr2 \tr \yi \yr\right)
-\qi1^2 \ti y \yr
+\qi1^2 \tr y \yi
-\qr1^2 \ti y \yr
+\qr1^2 \tr y \yi 
\left.\right]
\end{autobreak} \\[0.2cm]
\begin{autobreak}\MoveEqLeft
\ydiagram[*(lightgray)]{4}*[*(red)]{7,1}*[*(darkgreen)]{7,3}*[*(lightgray)]{7,7}~=~ \I \left[ \right.
-4  \left(q_3 \qr1 \ti \tr y\right.
  +\qi2 \qr2 t \ti \yi
  -\qi2 \qr2 t \tr \yr
  -\qi2 \qr2 \ti^2 y
  +\qi2 \qr2 \tr^2 y
  -\qi2^2 t \tr \yi
  +\qi2^2 \ti \tr y
  +\qr2^2 t \ti \yr
  \left.-\qr2^2 \ti \tr y\right)
-3 \left(q_3^2 t \ti \yr-q_3^2 t \tr \yi\right)
-2 \left(-q_3 \qi2 t^2 \yr\right.
  -q_3 \qi1 t \ti \yi
  +q_3 \qi2 t \tr y
  +q_3 \qi1 t \tr \yr
  +q_3 \qi1 \ti^2 y
  +q_3 \qi2 \ti^2 \yr
  -q_3 \qi2 \ti \tr \yi
  -q_3 \qi1 \tr^2 y
  +q_3 \qr2 t^2 \yi
  -q_3 \qr2 t \ti y
  -q_3 \qr1 t \ti \yr
  -q_3 \qr1 t \tr \yi
  +q_3 \qr2 \ti \tr \yr
  -q_3 \qr2 \tr^2 \yi
  -\qi1 \qr2 t^2 \yr
  +\qi2 \qr1 t^2 \yr
  -\qi2 \qr1 t \tr y
  +\qi1 \qr2 t \tr y
  -\qi2 \qr1 \ti^2 \yr
  +\qi1 \qr2 \ti^2 \yr
  -\qi1 \qr2 \ti \tr \yi
  +\qi2 \qr1 \ti \tr \yi
  +\qi1 \qi2 t^2 \yi
  -\qi1 \qi2 t \ti y
  +\qi1 \qi2 \ti \tr \yr
  -\qi1 \qi2 \tr^2 \yi
  +\qr1 \qr2 t^2 \yi
  -\qr1 \qr2 t \ti y
  +\qr1 \qr2 \ti \tr \yr
  \left.-\qr1 \qr2 \tr^2 \yi\right)
+\qi1^2 t \ti \yr
-\qi1^2 t \tr \yi
+\qr1^2 t \ti \yr
-\qr1^2 t \tr \yi
\left.\right]
\end{autobreak} \\[0.2cm]
\begin{autobreak}\MoveEqLeft
\ydiagram[*(lightgray)]{9,3}*[*(darkgreen)]{9,9}~=~ \I \left[\right. 
18\, y \yi \yr \qr1 q_3^2
-16 \left(\yi^2 \yr \qi2 \qr2^2\right.
  \left.-\yi \yr^2 \qi2^2 \qr2\right)
-12 \left(\yi q_3 \qi1 \qi2 y^2\right.
  +\yr q_3 \qi2 \qr1 y^2
  -\yr q_3 \qi1 \qr2 y^2
  +\yi q_3 \qr1 \qr2 y^2
  -\yi \yr q_3 \qi2^2 y
  +\yi \yr q_3 \qr2^2 y
  +\yi^2 q_3 \qi2 \qr2 y
  -\yr^2 q_3 \qi2 \qr2 y
  +\yi^2 \qi2 \qr1 \qr2 y
  +\yr^2 \qi2 \qr1 \qr2 y
  -\yi^2 \yr q_3 \qi2 \qr1
  \left.-\yi \yr^2 q_3 \qr1 \qr2\right)
+9 \left(y \yi^2 q_3^2 \qi1\right.
  \left.-y \yr^2 q_3^2 \qi1\right)
+8 \left(\qi2 \qr1 \qr2 y^3\right.
  -\yr \qi2^3 y^2
  +\yi \qr2^3 y^2
  -\yr \qi2 \qr2^2 y^2
  +\yi \qi2^2 \qr2 y^2
  -\yr^2 \qi1 \qi2^2 y
  +\yi^2 \qi1 \qr2^2 y
  +\yi^2 \yr \qi2^3
  -\yi \yr^2 \qr2^3
  +\yr^3 \qi2 \qr2^2
  \left.-\yi^3 \qi2^2 \qr2\right)
+6 \left(q_3 \qi1 \qi2 \yi^3\right.
  +\yr q_3 \qi1 \qr2 \yi^2
  +\yr \qi1 \qr1 \qr2 \yi^2
  -\yr^2 q_3 \qi1 \qi2 \yi
  +\yr^2 \qi1 \qi2 \qr1 \yi
  \left.-\yr^3 q_3 \qi1 \qr2\right)
+4 \left(\qi1 \qi2^2 y^3\right.
  -\qi1 \qr2^2 y^3
  +\yr \qi2 \qr1^2 y^2
  +\yr \qi1^2 \qi2 y^2
  -\yi \qi1^2 \qr2 y^2
  -\yi \qr1^2 \qr2 y^2
  -\yi^2 \qi1 \qi2^2 y
  +\yr^2 \qi1 \qr2^2 y
  +\yi \yr \qr1 \qr2^2 y
  +\yi \yr \qi2^2 \qr1 y
  -\yi^2 \yr \qi2 \qr1^2
  \left.+\yi \yr^2 \qr1^2 \qr2\right)
-2 \left(\qi1 \qi2 \qr1 \yi^3\right.
  -\qi1^2 \qr2 \yi^3
  -\yr \qi1^2 \qi2 \yi^2
  +y \yr \qr1^3 \yi
  +y \yr \qi1^2 \qr1 \yi
  +\yr^2 \qi1^2 \qr2 \yi
  +\yr^3 \qi1^2 \qi2
  \left.+\yr^3 \qi1 \qr1 \qr2\right)
-y \yi^2 \qi1^3
+y \yr^2 \qi1^3
-y \yi^2 \qr1^2 \qi1
+y \yr^2 \qr1^2 \qi1
\left.\right]
\end{autobreak} \\[0.2cm]
\begin{autobreak}\MoveEqLeft
\ydiagram[*(lightgray)]{9,3}*[*(red)]{9,9}~=~ \I \left[\right.
18\, t \ti \tr \qr1 q_3^2
-16 \left(\ti^2 \tr \qi2 \qr2^2\right.
  \left.-\ti \tr^2 \qi2^2 \qr2\right)
-12 \left(\ti q_3 \qi1 \qi2 t^2\right.
  +\tr q_3 \qi2 \qr1 t^2
  -\tr q_3 \qi1 \qr2 t^2
  +\ti q_3 \qr1 \qr2 t^2
  -\ti \tr q_3 \qi2^2 t
  +\ti \tr q_3 \qr2^2 t
  +\ti^2 q_3 \qi2 \qr2 t
  -\tr^2 q_3 \qi2 \qr2 t
  +\ti^2 \qi2 \qr1 \qr2 t
  +\tr^2 \qi2 \qr1 \qr2 t
  -\ti^2 \tr q_3 \qi2 \qr1
  \left.-\ti \tr^2 q_3 \qr1 \qr2\right)
+9 \left(t \ti^2 q_3^2 \qi1\right.
  \left.-t \tr^2 q_3^2 \qi1\right)
+8 \left(\qi2 \qr1 \qr2 t^3\right.
  -\tr \qi2^3 t^2
  +\ti \qr2^3 t^2
  -\tr \qi2 \qr2^2 t^2
  +\ti \qi2^2 \qr2 t^2
  -\tr^2 \qi1 \qi2^2 t
  +\ti^2 \qi1 \qr2^2 t
  +\ti^2 \tr \qi2^3
  -\ti \tr^2 \qr2^3
  +\tr^3 \qi2 \qr2^2
  \left.-\ti^3 \qi2^2 \qr2\right)
+6 \left(q_3 \qi1 \qi2 \ti^3\right.
  +\tr q_3 \qi1 \qr2 \ti^2
  +\tr \qi1 \qr1 \qr2 \ti^2
  -\tr^2 q_3 \qi1 \qi2 \ti
  +\tr^2 \qi1 \qi2 \qr1 \ti
  \left.-\tr^3 q_3 \qi1 \qr2\right)
+4 \left(\qi1 \qi2^2 t^3\right.
  -\qi1 \qr2^2 t^3
  +\tr \qi2 \qr1^2 t^2
  +\tr \qi1^2 \qi2 t^2
  -\ti \qi1^2 \qr2 t^2
  -\ti \qr1^2 \qr2 t^2
  -\ti^2 \qi1 \qi2^2 t
  +\tr^2 \qi1 \qr2^2 t
  +\ti \tr \qr1 \qr2^2 t
  +\ti \tr \qi2^2 \qr1 t
  -\ti^2 \tr \qi2 \qr1^2
  \left.+\ti \tr^2 \qr1^2 \qr2\right)
-2 \left(\qi1 \qi2 \qr1 \ti^3\right.
  -\qi1^2 \qr2 \ti^3
  -\tr \qi1^2 \qi2 \ti^2
  +t \tr \qr1^3 \ti
  +t \tr \qi1^2 \qr1 \ti
  +\tr^2 \qi1^2 \qr2 \ti
  +\tr^3 \qi1^2 \qi2
  \left.+\tr^3 \qi1 \qr1 \qr2\right)
-t \ti^2 \qi1^3
+t \tr^2 \qi1^3
-t \ti^2 \qr1^2 \qi1
+t \tr^2 \qr1^2 \qi1
\left.\right]
\end{autobreak} \\[0.2cm]
\begin{autobreak}\MoveEqLeft
\ydiagram[*(lightgray)]{9,3}*[*(darkgreen)]{9,7}*[*(red)]{9,9}~=~ \I \left[\right. 
-32 \left(\ti \yi \yr \qi2 \qr2^2\right.
  \left.-\tr \yi \yr \qi2^2 \qr2\right)
+24 \left(t \qi2 \qr1 \qr2 y^2\right.
  -t \yi q_3 \qi1 \qi2 y
  -t \yr q_3 \qi2 \qr1 y
  +t \yr q_3 \qi1 \qr2 y
  -\ti \yi q_3 \qi2 \qr2 y
  +\tr \yr q_3 \qi2 \qr2 y
  -t \yi q_3 \qr1 \qr2 y
  -\ti \yi \qi2 \qr1 \qr2 y
  -\tr \yr \qi2 \qr1 \qr2 y
  +\tr \yr^2 \qi2 \qr2^2
  +\ti \yi \yr q_3 \qi2 \qr1
  -\ti \yi^2 \qi2^2 \qr2
  \left.+\tr \yi \yr q_3 \qr1 \qr2\right)
+18 \left(\ti q_3 \qi1 \qi2 \yi^2\right.
  +\ti y q_3^2 \qi1 \yi
  +\tr y q_3^2 \qr1 \yi
  +t \yr q_3^2 \qr1 \yi
  -\tr y \yr q_3^2 \qi1
  +\ti y \yr q_3^2 \qr1
  \left.-\tr \yr^2 q_3 \qi1 \qr2\right)
+16 \left(-t y \yr \qi2^3\right.
  +\ti \yi \yr \qi2^3
  -\tr y \yr \qi1 \qi2^2
  +\ti \yr^2 \qr2 \qi2^2
  +t y \yi \qr2 \qi2^2
  -\tr \yi^2 \qr2^2 \qi2
  -t y \yr \qr2^2 \qi2
  +t y \yi \qr2^3
  -\tr \yi \yr \qr2^3
  \left.+\ti y \yi \qi1 \qr2^2\right)
+12 \left(t \qi1 \qi2^2 y^2\right.
  -t \qi1 \qr2^2 y^2
  -\ti q_3 \qi1 \qi2 y^2
  -\tr q_3 \qi2 \qr1 y^2
  +\tr q_3 \qi1 \qr2 y^2
  -\ti q_3 \qr1 \qr2 y^2
  +\tr \yi q_3 \qi2^2 y
  +\ti \yr q_3 \qi2^2 y
  -\tr \yi q_3 \qr2^2 y
  -\ti \yr q_3 \qr2^2 y
  +t \yi \yr q_3 \qi2^2
  -t \yi \yr q_3 \qr2^2
  -\tr \yi \yr q_3 \qi1 \qi2
  +\tr \yi^2 q_3 \qi2 \qr1
  +\tr \yi \yr \qi1 \qi2 \qr1
  +\ti \yi \yr q_3 \qi1 \qr2
  -t \yi^2 q_3 \qi2 \qr2
  +t \yr^2 q_3 \qi2 \qr2
  +\ti \yr^2 q_3 \qr1 \qr2
  +\ti \yi \yr \qi1 \qr1 \qr2
  -t \yi^2 \qi2 \qr1 \qr2
  \left.-t \yr^2 \qi2 \qr1 \qr2\right)
+9 \left(t \yi^2 q_3^2 \qi1\right.
  \left.-t \yr^2 q_3^2 \qi1\right)
-8 \left(\tr y^2 \qi2^3\right.
  -\tr \yi^2 \qi2^3
  +t \yr^2 \qi1 \qi2^2
  +\ti y \yi \qi1 \qi2^2
  -\ti y^2 \qr2 \qi2^2
  -t y \yr \qi1^2 \qi2
  -t y \yr \qr1^2 \qi2
  +\ti \yi \yr \qr1^2 \qi2
  +\tr y^2 \qr2^2 \qi2
  -\ti y^2 \qr2^3
  +\ti \yr^2 \qr2^3
  -t \yi^2 \qi1 \qr2^2
  -\tr y \yr \qi1 \qr2^2
  +t y \yi \qi1^2 \qr2
  +t y \yi \qr1^2 \qr2
  \left.-\tr \yi \yr \qr1^2 \qr2\right)
+6 \left(-\ti \qi1 \qi2 \qr1 \yi^2\right.
  +\ti \qi1^2 \qr2 \yi^2
  +\tr q_3 \qi1 \qr2 \yi^2
  +\tr \qi1 \qr1 \qr2 \yi^2
  -\tr \yr^2 \qi1^2 \qi2
  -\ti \yr^2 q_3 \qi1 \qi2
  +\ti \yr^2 \qi1 \qi2 \qr1
  \left.-\tr \yr^2 \qi1 \qr1 \qr2\right)
-4 \left(-\tr y^2 \qi2 \qi1^2\right.
  -\ti \yi \yr \qi2 \qi1^2
  +\ti y^2 \qr2 \qi1^2
  +\tr \yi \yr \qr2 \qi1^2
  +t \yi^2 \qi2^2 \qi1
  -t \yr^2 \qr2^2 \qi1
  -\tr y^2 \qi2 \qr1^2
  +\tr \yi^2 \qi2 \qr1^2
  -\tr y \yi \qr1 \qr2^2
  -\ti y \yr \qr1 \qr2^2
  -t \yi \yr \qr1 \qr2^2
  -\tr y \yi \qi2^2 \qr1
  -\ti y \yr \qi2^2 \qr1
  -t \yi \yr \qi2^2 \qr1
  +\ti y^2 \qr1^2 \qr2
  \left.-\ti \yr^2 \qr1^2 \qr2\right)
-2 \left(\ti y \yi \qi1^3\right.
  -\tr y \yr \qi1^3
  -\tr \yi^2 \qi2 \qi1^2
  +\tr y \yi \qr1 \qi1^2
  +\ti y \yr \qr1 \qi1^2
  +t \yi \yr \qr1 \qi1^2
  +\ti \yr^2 \qr2 \qi1^2
  +\ti y \yi \qr1^2 \qi1
  -\tr y \yr \qr1^2 \qi1
  +\tr y \yi \qr1^3
  +\ti y \yr \qr1^3
  \left.+t \yi \yr \qr1^3\right)
-t \yi^2 \qi1^3
+t \yr^2 \qi1^3
-t \yi^2 \qr1^2 \qi1
+t \yr^2 \qr1^2 \qi1
\left.\right]
\end{autobreak} \\[0.2cm]
\begin{autobreak}\MoveEqLeft
\ydiagram[*(lightgray)]{9,3}*[*(red)]{9,7}*[*(darkgreen)]{9,9}~=~ \I \left[\right.
-32 \left(\ti \tr \yi \qi2 \qr2^2\right.
  \left.-\ti \tr \yr \qi2^2 \qr2\right)
+24 \left(y \qi2 \qr1 \qr2 t^2\right.
  -\ti y q_3 \qi1 \qi2 t
  -\tr y q_3 \qi2 \qr1 t
  +\tr y q_3 \qi1 \qr2 t
  -\ti \yi q_3 \qi2 \qr2 t
  +\tr \yr q_3 \qi2 \qr2 t
  -\ti y q_3 \qr1 \qr2 t
  -\ti \yi \qi2 \qr1 \qr2 t
  -\tr \yr \qi2 \qr1 \qr2 t
  +\tr^2 \yr \qi2 \qr2^2
  +\ti \tr \yi q_3 \qi2 \qr1
  -\ti^2 \yi \qi2^2 \qr2
  \left.+\ti \tr \yr q_3 \qr1 \qr2\right)
+18 \left(\yi q_3 \qi1 \qi2 \ti^2\right.
  +t \yi q_3^2 \qi1 \ti
  +\tr y q_3^2 \qr1 \ti
  +t \yr q_3^2 \qr1 \ti
  -t \tr \yr q_3^2 \qi1
  +t \tr \yi q_3^2 \qr1
  \left.-\tr^2 \yr q_3 \qi1 \qr2\right)
+16 \left(-t \tr y \qi2^3\right.
  +\ti \tr \yi \qi2^3
  -t \tr \yr \qi1 \qi2^2
  +t \ti y \qr2 \qi2^2
  +\tr^2 \yi \qr2 \qi2^2
  -t \tr y \qr2^2 \qi2
  -\ti^2 \yr \qr2^2 \qi2
  +t \ti y \qr2^3
  -\ti \tr \yr \qr2^3
  \left.+t \ti \yi \qi1 \qr2^2\right)
+12 \left(y \qi1 \qi2^2 t^2\right.
  -y \qi1 \qr2^2 t^2
  -\yi q_3 \qi1 \qi2 t^2
  -\yr q_3 \qi2 \qr1 t^2
  +\yr q_3 \qi1 \qr2 t^2
  -\yi q_3 \qr1 \qr2 t^2
  +\tr \yi q_3 \qi2^2 t
  +\ti \yr q_3 \qi2^2 t
  -\tr \yi q_3 \qr2^2 t
  -\ti \yr q_3 \qr2^2 t
  +\ti \tr y q_3 \qi2^2
  -\ti \tr y q_3 \qr2^2
  -\ti \tr \yr q_3 \qi1 \qi2
  +\ti^2 \yr q_3 \qi2 \qr1
  +\ti \tr \yr \qi1 \qi2 \qr1
  +\ti \tr \yi q_3 \qi1 \qr2
  -\ti^2 y q_3 \qi2 \qr2
  +\tr^2 y q_3 \qi2 \qr2
  +\tr^2 \yi q_3 \qr1 \qr2
  +\ti \tr \yi \qi1 \qr1 \qr2
  -\ti^2 y \qi2 \qr1 \qr2
  \left.-\tr^2 y \qi2 \qr1 \qr2\right)
+9 \left(\ti^2 y q_3^2 \qi1\right.
  \left.-\tr^2 y q_3^2 \qi1\right)
+8 \left(\ti^2 \yr \qi2^3\right.
  -t^2 \yr \qi2^3
  -\tr^2 y \qi1 \qi2^2
  -t \ti \yi \qi1 \qi2^2
  +t^2 \yi \qr2 \qi2^2
  +t \tr y \qi1^2 \qi2
  +t \tr y \qr1^2 \qi2
  -\ti \tr \yi \qr1^2 \qi2
  -t^2 \yr \qr2^2 \qi2
  +t^2 \yi \qr2^3
  -\tr^2 \yi \qr2^3
  +\ti^2 y \qi1 \qr2^2
  +t \tr \yr \qi1 \qr2^2
  -t \ti y \qi1^2 \qr2
  -t \ti y \qr1^2 \qr2
  \left.+\ti \tr \yr \qr1^2 \qr2\right)
+6 \left(-\yi \qi1 \qi2 \qr1 \ti^2\right.
  +\yi \qi1^2 \qr2 \ti^2
  +\yr q_3 \qi1 \qr2 \ti^2
  +\yr \qi1 \qr1 \qr2 \ti^2
  -\tr^2 \yr \qi1^2 \qi2
  -\tr^2 \yi q_3 \qi1 \qi2
  +\tr^2 \yi \qi1 \qi2 \qr1
  \left.-\tr^2 \yr \qi1 \qr1 \qr2\right)
-4 \left(-\ti \tr \yi \qi2 \qi1^2\right.
  -t^2 \yr \qi2 \qi1^2
  +t^2 \yi \qr2 \qi1^2
  +\ti \tr \yr \qr2 \qi1^2
  +\ti^2 y \qi2^2 \qi1
  -\tr^2 y \qr2^2 \qi1
  -t^2 \yr \qi2 \qr1^2
  +\ti^2 \yr \qi2 \qr1^2
  -\ti \tr y \qr1 \qr2^2
  -t \tr \yi \qr1 \qr2^2
  -t \ti \yr \qr1 \qr2^2
  -\ti \tr y \qi2^2 \qr1
  -t \tr \yi \qi2^2 \qr1
  -t \ti \yr \qi2^2 \qr1
  +t^2 \yi \qr1^2 \qr2
  \left.-\tr^2 \yi \qr1^2 \qr2\right)
-2 \left(t \ti \yi \qi1^3\right.
  -t \tr \yr \qi1^3
  -\ti^2 \yr \qi2 \qi1^2
  +\ti \tr y \qr1 \qi1^2
  +t \tr \yi \qr1 \qi1^2
  +t \ti \yr \qr1 \qi1^2
  +\tr^2 \yi \qr2 \qi1^2
  +t \ti \yi \qr1^2 \qi1
  -t \tr \yr \qr1^2 \qi1
  +\ti \tr y \qr1^3
  +t \tr \yi \qr1^3
  \left.+t \ti \yr \qr1^3\right)
-\ti^2 y \qi1^3
+\tr^2 y \qi1^3
-\ti^2 y \qr1^2 \qi1
+\tr^2 y \qr1^2 \qi1
\left.\right]
\end{autobreak}
\end{align}

\pagebreak
\normalsize

\section{Syzygies}\label{app:syzygies}
Here we list syzygies that we have explicitly constructed in the course of this work.
\renewcommand{\arraystretch}{1.05}
\begin{table}[!h]
\caption{\label{tab:Syzygies} List of the syzygies of the 2HDM scalar sector.
The first columns give the order, structure and CP transformation behavior of the relation. 
The following columns summarize the corresponding coefficients of the multi-graded Hilbert series (HS), 
Plethystic logarithm (PL) as well as the number of new relations 
that are arising (as evidence that the PL coefficient indeed gives this number correctly). 
Furthermore, we list the total number of power products of generating invariants which
give rise to the same structure (this is an indicator of how many 
previously found ``old'' relations appear in addition to possible new relations).
The last column gives important structures of invariants that appear in the 
given syzygy.}
\vspace{0.2cm}
\centerline{
\begin{tabular}{ccccccccc}
\hline
\hline
  order  & structure & CP & HS & PL & PP & new rels. & old rels. & comments \\
\hline
$6$  & $q^2y^2t^2$ & $+$ & $6$ & $-1$ & $7$ & $1$ & - & $\Inv{1,1,1}^2$, \Eqref{eq:Syz6} \\ 
$7$  & $q^3y^2t^2$ & $+$ & $7$ & $-1$ & $8$ & $1$ & - & \\
     & $q^2y^3t^2$ / $q^2y^2t^3$  & $-$ & $3$ & $-1$ & $4$ & $1$ / $1$ & - & \Eqref{eq:CPoddSyz7} \\
$8$  & $q^2y^4t^2$ / $q^2y^2t^4$ & $+$ & $9$ & $-1$ & $11$ & $1$ / $1$ & $1$ / $1$ & $\Jnv{1,2,1}^2$ / $\Jnv{1,1,2}^2$ \\
     & $q^2y^3t^3$ & $+$ & $9$ & $-1$ & $10$ & $1$ & - & $\Jnv{1,2,1}\times\Jnv{1,1,2}$ \\ 
     & $q^4y^2t^2$ & $+$ & $10$ & $-1$ & $12$ & $1$ & $1$ & $\Inv{2,1,1}^2-\left(\Inv{2,2,0}\times\Inv{2,0,2}\right)$ \\
     & $q^3y^4t$ / $q^3yt^4$ &   $-$ & $4$ & $-1$ & $5$ & $1$ / $1$ & - & $\Inv{2,2,0}\Jnv{1,2,1}$ / $\Inv{2,0,2}\Jnv{1,1,2}$, \Eqref{eq:CPoddSyz8} \\
     & $q^3y^3t^2$ /$q^3y^2t^3$ & $-$ & $6$ & $-3$ & $9$ & $3$ / $3$ & - & $\Inv{2,2,0}\Jnv{1,1,2}$ / $\Inv{2,0,2}\Jnv{1,2,1}$   \\
     &  & & & & & &                                                      & $\Inv{2,1,1}\Jnv{1,2,1}$ / $\Inv{2,1,1}\Jnv{1,1,2}$   \\
$9$  & $q^3y^3t^3$ & $+$ & $13$ & $-2$ & $17$ & $2$ & $2$ & \\
     & $q^3y^4t^2$ / $q^3y^2t^4$ & $+$ & $12$ & $-1$ & $15$ & $1$ / $1$ & $2$ / $2$ & \\
     & $q^4y^4t$ / $q^4yt^4$ & $-$ & $5$ & $-1$ & $6$ & $1$ / $1$ & - & \\
     & $q^4y^3t^2$ / $q^4y^2t^3$ & $-$ & $7$ & $-3$ & $11$ & $3$ / $3$ & $1$ / $1$ & \\
     & $\vdots$ & & & $\vdots$ & & $\vdots$ & $\vdots$ \\
$10$ & $q^4y^4t^2$ / $q^4y^2t^4$ & $+$ & $18$ & $-3$ & $22$ & $3$ / $3$ & $1$ / $1$ & $\Jnv{2,2,1}^2$ / $\Jnv{2,1,2}^2$, \Eqref{eq:CPevenSyz10} \\
     & $q^5yt^4$ / $q^5y^4t$     & $-$ & $7$ & $-1$ & $9$ & $1$ / $1$ & $1$ / $1$ &   \\
     & $q^5y^3t^2$ / $q^5y^2t^3$ & $-$ & $10$ & $-2$ & $13$ & $2$ / $2$ & $1$ / $1$ &    \\
     & $\vdots$ & & & $\vdots$ & & $\vdots$ & $\vdots$ \\
$12$ & $q^6y^6$ / $q^6t^6$ & $+$ & $10$ & $-1$ & $11$ & $1$ / $1$ & - & $\Jnv{3,3,0}^2$ / $\Jnv{3,0,3}^2$ \\
     & $q^6y^4t^2$ / $q^6y^2t^4$ & $+$ & $27$ & $-2$ & $38$ & $2$ / $2$ & $9$ / $9$ & $\Jnv{3,2,1}^2$ / $\Jnv{3,1,2}^2$ \\
     & $q^6y^3t^3$ & $+$ & $28$ & $-2$ & $43$ & $2$ & $13$ & $\Jnv{3,3,0}\times\Jnv{3,0,3}$, $\Jnv{3,2,1}\times\Jnv{3,1,2}$ \\
     & $\vdots$ & & & $\vdots$ & & $\vdots$ & $\vdots$ \\
\hline
\hline
\end{tabular}}
\end{table}

\bibliography{Bibliography}

\providecommand{\bysame}{\leavevmode\hbox to3em{\hrulefill}\thinspace}
\frenchspacing
\newcommand{\origttfamily}{}
\let\origttfamily=\ttfamily
\renewcommand{\ttfamily}{\origttfamily \hyphenchar\font=`\-}

\begin{thebibliography}{10}

\bibitem{Branco:2011iw}
G.~C. Branco, P.~M. Ferreira, L.~Lavoura, M.~N. Rebelo, M.~Sher, and J.~P.
  Silva, Phys. Rept. \textbf{516} (2012), 1, \texttt{arXiv:1106.0034} [hep-ph].

\bibitem{Santamaria:1993ah}
A.~Santamaria, Phys. Lett. \textbf{B305} (1993), 90,
  \texttt{arXiv:hep-ph/9302301} [hep-ph].

\bibitem{Branco:1999fs}
G.~C. Branco, L.~Lavoura, and J.~P. Silva, Int. Ser. Monogr. Phys. \textbf{103}
  (1999), 1.

\bibitem{Jarlskog:1985ht}
C.~Jarlskog, Phys. Rev. Lett. \textbf{55} (1985), 1039.

\bibitem{Bernabeu:1986fc}
J.~Bernabeu, G.~C. Branco, and M.~Gronau, Phys. Lett. \textbf{169B} (1986),
  243.

\bibitem{Branco:1986gr}
G.~C. Branco, L.~Lavoura, and M.~N. Rebelo, Phys. Lett. \textbf{B180} (1986),
  264.

\bibitem{Botella:1994cs}
F.~J. Botella and J.~P. Silva, Phys. Rev. \textbf{D51} (1995), 3870,
  \texttt{arXiv:hep-ph/9411288} [hep-ph].

\bibitem{Lavoura:1994fv}
L.~Lavoura and J.~P. Silva, Phys. Rev. \textbf{D50} (1994), 4619,
  \texttt{arXiv:hep-ph/9404276} [hep-ph].

\bibitem{Gunion:2005ja}
J.~F. Gunion and H.~E. Haber, Phys. Rev. \textbf{D72} (2005), 095002,
  \texttt{arXiv:hep-ph/0506227} [hep-ph].

\bibitem{Branco:2005em}
G.~C. Branco, M.~N. Rebelo, and J.~I. Silva-Marcos, Phys. Lett. \textbf{B614}
  (2005), 187, \texttt{arXiv:hep-ph/0502118} [hep-ph].

\bibitem{Ivanov:2005hg}
I.~P. Ivanov, Phys. Lett. \textbf{B632} (2006), 360,
  \texttt{arXiv:hep-ph/0507132} [hep-ph].

\bibitem{Nishi:2006tg}
C.~C. Nishi, Phys. Rev. \textbf{D74} (2006), 036003,
  \texttt{arXiv:hep-ph/0605153} [hep-ph], [Erratum: Phys.
  Rev.D76,119901(2007)].

\bibitem{Lebedev:2002wq}
O.~Lebedev, Phys. Rev. \textbf{D67} (2003), 015013,
  \texttt{arXiv:hep-ph/0209023} [hep-ph].

\bibitem{Dreiner:2007yz}
H.~K. Dreiner, J.~S. Kim, O.~Lebedev, and M.~Thormeier, Phys. Rev. \textbf{D76}
  (2007), 015006, \texttt{arXiv:hep-ph/0703074} [HEP-PH].

\bibitem{Ivanov:2015mwl}
I.~P. Ivanov and J.~P. Silva, Phys. Rev. \textbf{D93} (2016), no.~9, 095014,
  \texttt{arXiv:1512.09276} [hep-ph].

\bibitem{Haber:2018iwr}
H.~E. Haber, O.~M. Ogreid, P.~Osland, and M.~N. Rebelo,
  \texttt{arXiv:1808.08629} [hep-ph].

\bibitem{Ivanov:2018ime}
I.~P. Ivanov, C.~C. Nishi, J.~o.~P. Silva, and A.~Trautner,
  \texttt{arXiv:1810.13396} [hep-ph].

\bibitem{Jenkins:2007ip}
E.~E. Jenkins and A.~V. Manohar, Nucl. Phys. \textbf{B792} (2008), 187,
  \texttt{arXiv:0706.4313} [hep-ph].

\bibitem{Jenkins:2009dy}
E.~E. Jenkins and A.~V. Manohar, JHEP \textbf{10} (2009), 094,
  \texttt{arXiv:0907.4763} [hep-ph].

\bibitem{Hanany:2010vu}
A.~Hanany, E.~E. Jenkins, A.~V. Manohar, and G.~Torri, JHEP \textbf{03} (2011),
  096, \texttt{arXiv:1010.3161} [hep-ph].

\bibitem{Davidson:2005cw}
S.~Davidson and H.~E. Haber, Phys. Rev. \textbf{D72} (2005), 035004,
  \texttt{arXiv:hep-ph/0504050} [hep-ph], [Erratum: Phys.
  Rev.D72,099902(2005)].

\bibitem{Haber:2006ue}
H.~E. Haber and D.~O'Neil, Phys. Rev. \textbf{D74} (2006), 015018,
  \texttt{arXiv:hep-ph/0602242} [hep-ph], [Erratum: Phys.
  Rev.D74,no.5,059905(2006)].

\bibitem{Grzadkowski:2016szj}
B.~Grzadkowski, O.~M. Ogreid, and P.~Osland, Phys. Rev. \textbf{D94} (2016),
  no.~11, 115002, \texttt{arXiv:1609.04764} [hep-ph].

\bibitem{Bento:2017eti}
M.~P. Bento, H.~E. Haber, J.~C. Romão, and J.~P. Silva, JHEP \textbf{11}
  (2017), 095, \texttt{arXiv:1708.09408} [hep-ph].

\bibitem{Ogreid:2018bjq}
O.~M. Ogreid, PoS \textbf{CORFU2017} (2018), 065, \texttt{arXiv:1803.09351}
  [hep-ph].

\bibitem{Feldmann:2015nia}
T.~Feldmann, T.~Mannel, and S.~Schwertfeger, JHEP \textbf{10} (2015), 007,
  \texttt{arXiv:1507.00328} [hep-ph].

\bibitem{Chiu:2015ega}
S.~H. Chiu and T.~K. Kuo, Phys. Lett. \textbf{B760} (2016), 544,
  \texttt{arXiv:1510.07368} [hep-ph].

\bibitem{Chiu:2016qra}
S.~H. Chiu and T.~K. Kuo, Phys. Rev. \textbf{D93} (2016), no.~9, 093006,
  \texttt{arXiv:1603.04568} [hep-ph].

\bibitem{Herren:2017uxn}
F.~Herren, L.~Mihaila, and M.~Steinhauser, Phys. Rev. \textbf{D97} (2018),
  no.~1, 015016, \texttt{arXiv:1712.06614} [hep-ph].

\bibitem{Bednyakov:2018cmx}
A.~V. Bednyakov, \texttt{arXiv:1809.04527} [hep-ph].

\bibitem{Bijnens:2018rqw}
J.~Bijnens, J.~Oredsson, and J.~Rathsman, \texttt{arXiv:1810.04483} [hep-ph].

\bibitem{Varzielas:2016zjc}
I.~de~Medeiros~Varzielas, S.~F. King, C.~Luhn, and T.~Neder, Phys. Rev.
  \textbf{D94} (2016), no.~5, 056007, \texttt{arXiv:1603.06942} [hep-ph].

\bibitem{Berger:2018dxg}
D.~Berger, J.~N. Howard, and A.~Rajaraman, \texttt{arXiv:1806.04332} [hep-th].

\bibitem{stanley1979}
R.~P. Stanley, Bull. Amer. Math. Soc. (N.S.) \textbf{1} (1979), no.~3, 475.

\bibitem{Sturmfels:2008}
B.~Sturmfels, \emph{Algorithms in invariant theory (texts and monographs in
  symbolic computation)}, 2nd ed.; vii, 197 pp.; 5 figs. ed., Springer
  Publishing Company, Incorporated, 2008.

\bibitem{1994dg.ga.....8003G}
E.~{Getzler} and M.~M. {Kapranov}, \emph{{Modular operads}}, in \emph{eprint
  arXiv:dg-ga/9408003}, August 1994.

\bibitem{Labastida:2001ts}
J.~M.~F. Labastida and M.~Marino, \texttt{arXiv:math/0104180} [math-qa].

\bibitem{Benvenuti:2006qr}
S.~Benvenuti, B.~Feng, A.~Hanany, and Y.-H. He, JHEP \textbf{11} (2007), 050,
  \texttt{arXiv:hep-th/0608050} [hep-th].

\bibitem{Feng:2007ur}
B.~Feng, A.~Hanany, and Y.-H. He, JHEP \textbf{03} (2007), 090,
  \texttt{arXiv:hep-th/0701063} [hep-th].

\bibitem{Noma:2006pe}
Y.~Noma, T.~Nakatsu, and T.~Tamakoshi, \texttt{arXiv:hep-th/0611324} [hep-th].

\bibitem{Butti:2007jv}
A.~Butti, D.~Forcella, A.~Hanany, D.~Vegh, and A.~Zaffaroni, JHEP \textbf{11}
  (2007), 092, \texttt{arXiv:0705.2771} [hep-th].

\bibitem{Gray:2008yu}
J.~Gray, A.~Hanany, Y.-H. He, V.~Jejjala, and N.~Mekareeya, JHEP \textbf{05}
  (2008), 099, \texttt{arXiv:0803.4257} [hep-th].

\bibitem{Hanany:2008kn}
A.~Hanany and N.~Mekareeya, JHEP \textbf{10} (2008), 012,
  \texttt{arXiv:0805.3728} [hep-th].

\bibitem{Hanany:2008sb}
A.~Hanany, N.~Mekareeya, and G.~Torri, Nucl. Phys. \textbf{B825} (2010), 52,
  \texttt{arXiv:0812.2315} [hep-th].

\bibitem{Hanany:2014dia}
A.~Hanany and R.~Kalveks, JHEP \textbf{10} (2014), 152,
  \texttt{arXiv:1408.4690} [hep-th].

\bibitem{Bourget:2017tmt}
A.~Bourget and A.~Pini, JHEP \textbf{10} (2017), 033, \texttt{arXiv:1706.03781}
  [hep-th].

\bibitem{Henning:2015daa}
B.~Henning, X.~Lu, T.~Melia, and H.~Murayama, Commun. Math. Phys. \textbf{347}
  (2016), no.~2, 363, \texttt{arXiv:1507.07240} [hep-th].

\bibitem{Henning:2015alf}
B.~Henning, X.~Lu, T.~Melia, and H.~Murayama, JHEP \textbf{08} (2017), 016,
  \texttt{arXiv:1512.03433} [hep-ph].

\bibitem{Henning:2017fpj}
B.~Henning, X.~Lu, T.~Melia, and H.~Murayama, JHEP \textbf{10} (2017), 199,
  \texttt{arXiv:1706.08520} [hep-th].

\bibitem{Lehman:2015via}
L.~Lehman and A.~Martin, Phys. Rev. \textbf{D91} (2015), 105014,
  \texttt{arXiv:1503.07537} [hep-ph].

\bibitem{Fonseca:2011sy}
R.~M. Fonseca, Comput. Phys. Commun. \textbf{183} (2012), 2298,
  \texttt{arXiv:1106.5016} [hep-ph].

\bibitem{Keppeler:2013yla}
S.~Keppeler and M.~Sjödahl, J. Math. Phys. \textbf{55} (2014), 021702,
  \texttt{arXiv:1307.6147} [math-ph].

\bibitem{Alckock-Zeilinger:2016bss}
J.~Alcock-Zeilinger and H.~Weigert, J. Math. Phys. \textbf{58} (2017), no.~5,
  051701, \texttt{arXiv:1610.08801} [math-ph].

\bibitem{Alcock-Zeilinger:2016cva}
J.~Alcock-Zeilinger and H.~Weigert, J. Math. Phys. \textbf{58} (2017), no.~5,
  051703, \texttt{arXiv:1610.08802} [math-ph].

\bibitem{Alcock-Zeilinger:2016sxc}
J.~Alcock-Zeilinger and H.~Weigert, J. Math. Phys. \textbf{58} (2017), no.~5,
  051702, \texttt{arXiv:1610.10088} [math-ph].

\bibitem{Cvitanovic:1976am}
P.~Cvitanovic, Phys. Rev. \textbf{D14} (1976), 1536.

\bibitem{Cvitanovic:2008zz}
P.~Cvitanovic, \emph{{Group theory: Birdtracks, Lie's and exceptional groups}},
  2008.

\bibitem{Keppeler:2017kwt}
S.~Keppeler, \texttt{arXiv:1707.07280} [math-ph].

\bibitem{Noether:1916}
E.~Noether, Math. Ann. \textbf{77} (1916), 89–92.

\bibitem{Hochster:1974a}
M.~Hochster and J.~L. Roberts, Bulletin of the American Mathematical Society
  \textbf{80} (1974), no.~2, 281.

\bibitem{Hochster:1974b}
M.~Hochster and J.~L. Roberts, Advances in Mathematics \textbf{13} (1974), 115.

\bibitem{Knop1987}
F.~Knop and P.~Littelmann, Mathematische Zeitschrift \textbf{196} (1987),
  no.~2, 211.

\bibitem{KNOP198940}
F.~Knop, Journal of Algebra \textbf{127} (1989), no.~1, 40 .

\bibitem{Dreiner:2008tw}
H.~K. Dreiner, H.~E. Haber, and S.~P. Martin, Phys. Rept. \textbf{494} (2010),
  1, \texttt{arXiv:0812.1594} [hep-ph].

\bibitem{Ecker:1987qp}
G.~Ecker, W.~Grimus, and H.~Neufeld, J. Phys. \textbf{A20} (1987), L807.

\bibitem{Bourget:2018ond}
A.~Bourget, A.~Pini, and D.~Rodríguez-Gómez, \texttt{arXiv:1804.01108}
  [hep-th].

\bibitem{2008arXiv0812.3082T}
N.~M. {Thi{\'e}ry}, ArXiv e-prints (2008), \texttt{arXiv:0812.3082} [math.CO].

\bibitem{2007math......1270K}
S.~A. {King}, ArXiv Mathematics e-prints (2007), \texttt{math/0701270}.

\bibitem{Nagel:2004sw}
F.~Nagel, \emph{{New aspects of gauge-boson couplings and the Higgs sector}},
  Ph.D. thesis, Heidelberg U., 2004.

\bibitem{Maniatis:2006fs}
M.~Maniatis, A.~von Manteuffel, O.~Nachtmann, and F.~Nagel, Eur. Phys. J.
  \textbf{C48} (2006), 805, \texttt{arXiv:hep-ph/0605184} [hep-ph].

\bibitem{Ivanov:2006yq}
I.~P. Ivanov, Phys. Rev. \textbf{D75} (2007), 035001,
  \texttt{arXiv:hep-ph/0609018} [hep-ph], [Erratum: Phys.
  Rev.D76,039902(2007)].

\bibitem{Maniatis:2007vn}
M.~Maniatis, A.~von Manteuffel, and O.~Nachtmann, Eur. Phys. J. \textbf{C57}
  (2008), 719, \texttt{arXiv:0707.3344} [hep-ph].

\bibitem{Ivanov:2007de}
I.~P. Ivanov, Phys. Rev. \textbf{D77} (2008), 015017, \texttt{arXiv:0710.3490}
  [hep-ph].

\bibitem{Ferreira:2010hy}
P.~M. Ferreira, M.~Maniatis, O.~Nachtmann, and J.~P. Silva, JHEP \textbf{08}
  (2010), 125, \texttt{arXiv:1004.3207} [hep-ph].

\bibitem{Ferreira:2010yh}
P.~M. Ferreira, H.~E. Haber, M.~Maniatis, O.~Nachtmann, and J.~P. Silva, Int.
  J. Mod. Phys. \textbf{A26} (2011), 769, \texttt{arXiv:1010.0935} [hep-ph].

\bibitem{Maniatis:2014oza}
M.~Maniatis and O.~Nachtmann, JHEP \textbf{02} (2015), 058,
  \texttt{arXiv:1408.6833} [hep-ph], [Erratum: JHEP10,149(2015)].

\bibitem{Ivanov:2014doa}
I.~P. Ivanov and C.~C. Nishi, JHEP \textbf{01} (2015), 021,
  \texttt{arXiv:1410.6139} [hep-ph].

\bibitem{Ehrenborg:1993}
R.~Ehrenborg and G.-C. Rota, European Journal of Combinatorics \textbf{14}
  (1993), no.~3, 157 .

\bibitem{Beecken:2011}
M.~Beecken, J.~Mittmann, and N.~Saxena, CoRR \textbf{abs/1102.2789} (2011),
  \texttt{1102.2789}.

\end{thebibliography}
\addcontentsline{toc}{section}{Bibliography}
\bibliographystyle{NewArXiv} 
\end{document}